# Towards a Switching-Algebraic Theory of Weighted Voting Systems: Exploring Restrictions on Coalition Formation

[1]**Ali Muhammad Rushdi** and [2,3]**Muhammad Ali Rushdi**

[1]Department of Electrical and Computer Engineering, Faculty of Engineering, King Abdulaziz University, P. O. Box 80200, Jeddah, 21589, Kingdom of Saudi Arabia
{arushdi@kau.edu.sa; arushdi@ieee.org}

[2]Department of Biomedical and Systems Engineering, Faculty of Engineering, Cairo University, Giza 12613, Arab Republic of Egypt
{mrushdi@eng1.cu.edu.eg}
[3]School of Information Technology, New Giza University, Giza 12256, Arab Republic of Egypt
{Muhammad.Rushdi@ngu.edu.eg}

**Abstract**: Mainstream voting theory concentrates on monotone voting systems, which comprise independent voters, and involve monotonically non-decreasing decision functions. This paper is a continuation of earlier efforts towards the development of a comprehensive switching-algebraic treatment for voting systems. The paper distinguishes itself from its predecessors by addressing the case when restrictions are imposed on the formation of coalitions. We explore the switching-algebraic computation of the Banzhaf indices for general and monotone or unrestricted systems. This computation is achieved via (a) two Boolean-quotient formulas that are valid when the voting system is not necessarily monotone (e.g., when coalition formation is restricted), (b) four Boolean differencing formulas and six Boolean-quotient formulas that are applicable when the decision switching function is a positively polarized unate one. We also provide switching-algebraic formulas for certain Banzhaf-related indices, including the power-to-initiate index (PII), and the power-to-prevent index (PPI), as well as satisfaction indices. Moreover, we briefly address other Banzhaf-related indices, including the Strict Power Index (SPI) and the Public Good Index (PGI). We illustrate the various indices formulas by way of four examples of voting systems, each considered first as an unrestricted monotone system and then subjected to a restriction on the formation of a coalition between two particular voters. In each of these examples, the restricted case involves (a) a loss of the original independence between the two variables representing the two restricted voters, and (b) a partial destruction of the original unateness of the decision function $f(\boldsymbol{X})$ of the voting system as this function is replaced by a constrained one $g(\boldsymbol{X})$ that ceases to be monoform in the two variables representing the restricted voters (though it remains monoform in the remaining variables). To handle the restricted case with switching-algebraic techniques, we (a) construct the constrained function $g(\boldsymbol{X})$ such that it is generally in agreement with the original function $f(\boldsymbol{X})$ except for the nullification of its Boolean quotient w.r.t. the two-literal product of the forbidden coalition, and (b) calculate the total Banzhaf power of each of the two restricted voters using only one of the two formulas that do not necessitate monotonicity of the decision function. We visualize the exact Universe of Discourse (probability sample space) for the restricted case as a Karnaugh-map-like structure, wherein the domain of the forbidden coalition is annihilated. For mathematical convenience, we use an actual Karnaugh map as a sample space, in which the domain of the forbidden coalition is restored, albeit with zero content. We show that, for a 2-out-of-3 system, the vector of total Banzhaf powers changes from a value of $\boldsymbol{TBP} = [2 \quad 2 \quad 2]^T$ to a value of $\boldsymbol{TBP} = [1 \quad 1 \quad 2]^T$ when the first two voters refuse to form a potentially valid coalition. We further generalize this system to an arbitrary k-out-of-n system, which serves as a general model for simple majority systems or super-majority systems, e.g., three-fifth, two-third, or three-quarter ones. We exemplify the arbitrary k-out-of-n system by a 5-out-of-8 one, visualized on an elegant and regular eight-variable Karnaugh map. Our results demonstrate that the Public Good Index (PGI) is exactly equal to the Total Banzhaf Power (TBP) for a general k-out-of-n system, not only when this system is unrestricted, but also when it is restricted through the lack of co-operation between two voting members. We present the reduced version of a scalar-weighted five-member voting system that nearly represents the Scottish Parliament of 2007. Here, the TBP and the PGI cease to be exactly the same, but they both indicate a loss of power for the two largest parties when they refuse to form a coalition, with the ironic rise of the third largest party to the status of the most powerful one. We check our results by repeated calculation via independent means, through exhaustive visualization of the entire sample space, or (when possible) through the reproduction of previously published results (albeit in a short-cut fashion). To make the paper self-contained, we provide an extensive introduction to the concept of a Boolean quotient, its general properties, its relations to the Boole-Shannon expansion and the Boolean difference, and its utility in interpreting various concepts of voting theory, especially those of voter desirability. The paper is hopefully of a significant pedagogical utility as it supplements the existing voting theory with an alternative perspective, an easier-to-comprehend methodology, and more handy and pictorial tools.

## 1. Introduction

The study of yes-no voting systems is dominantly a game-theoretic exploration [1-11], and it mainly concentrate on monotone systems [9, 12], which are ones whose decision functions $f(\boldsymbol{X})$ are positively polarized unate two-valued Boolean functions. Implicit in the definition of these systems is the assumption that system states or configurations are equally likely. This in turn necessitates that voters cast their votes independently of each other. Recently, several papers have strived to develop a supplemental switching-theoretic treatment for these systems [12-30]. The present paper follows in the footsteps of these papers, and it is essentially a sequel of [12], with a single notable innovation. This paper addresses the issue of incompatibility among voters (who cannot cooperate among themselves for ideological, political or socio-economic reasons) [31-42]. In particular, we target the situation where coalition formation is restricted, and hence the assumption of independence among certain voters is relaxed. This situation was handled earlier by Yakuba [33], who employed a two-stage procedure for evaluating the Banzhaf index using generating functions. In this paper, we choose not to use the efficient game-theoretic technique of generating functions [43], but rather rely mostly on the Boolean quotient concept of switching theory, a concept that we believe to be simpler and pedagogically more appealing.

Specifically, we mandate that two particular voters $X_m$ and $X_k$ never be in the same coalition (so that the two-literal product $X_m X_k$ cannot be part of any coalition), and hence we need to nullify any instance of the product $X_m X_k$ in the system decision function $f(\boldsymbol{X})$. Though we can attain this nullification in an *ad* hoc way (typically a simplified shortcut, See Section 6) we achieve it systematically by constructing the Boole-Shannon expansion w.r.t. the two variables that appear in the forbidden coalition. This Boole-Shannon expansion comprises $2^2 = 4$ Boolean quotients. We then convert the decision function $f(\boldsymbol{X})$ into a restricted one $g(\boldsymbol{X})$ by nullifying the Boolean quotient w.r.t the product $X_m X_k$, while leaving the remaining Boolean quotients intact. This nullification is a matter of mathematical convenience, as it has the same effect as that of nullifying the product $X_m X_k$ itself. Replacement of $f(\boldsymbol{X})$ by $g(\boldsymbol{X})$ destroys the unateness of $g(\boldsymbol{X})$ w.r.t. each of the two variables $X_m$ and $X_k$, but does not spoil its mono-polarization in the rest of the variables. This replacement also destroys the original total independence among voter variables as it causes a loss of the original independence between the two variables $X_m$ and $X_k$ representing the two restricted voters. We then calculate the total Banzhaf power of each of the two restricted voters $X_m$ and $X_k$ using only formulas that do not necessitate monotonicity of the decision function. We calculate the total Banzhaf power of each of the remaining variables using any convenient formula.

The organization of the rest of this paper is as follows. Section 2 deals with the switching-algebraic computation of the Banzhaf indices for general and monotone systems. This computation is achieved via (a) two Boolean-quotient formulas that are valid when the voting system is not necessarily monotone (e.g., when coalition formation is restricted), (b) four Boolean differencing formulas and six Boolean-quotient formulas that are applicable when the decision switching function is a positively polarized unate one. Section 2 also provides switching-algebraic formulas for certain Banzhaf-related indices, including the power-to-initiate index (PII), and the power-to-prevent index (PPI), as well as satisfaction indices. Finally, Section 2 also briefly addresses other Banzhaf-related indices, including the Strict Power Index (SPI) and the Public Good Index (PGI). Sections 3 to 6 illustrate the formulas of Section 2 by way of four examples of voting systems, each considered first as an unrestricted monotone system and then subjected to a restriction on the formation of a coalition between two particular voters. In each of these examples, the restricted case involves (a) a loss of the original independence between the two variables representing the two restricted voters, and (b) a partial destruction of the original unateness of the decision function $f(\boldsymbol{X})$ of the voting system as this function is replaced by a constrained one $g(\boldsymbol{X})$ that ceases to be monoform in the two variables representing the restricted voters (though it remains monoform in the remaining variables). To handle the restricted case with switching-algebraic techniques, we (a) construct the constrained function $g(\boldsymbol{X})$ such that it is generally in agreement with the original function $f(\boldsymbol{X})$ except for the nullification of its Boolean quotient w.r.t. the two-literal product of the forbidden coalition, and (b) calculate the total Banzhaf power of each of the two restricted voters using only one of the two formulas that do not necessitate monotonicity of the decision function. We visualize the exact Universe of Discourse (probability sample space) for the restricted case as a Karnaugh-map-like structure, wherein the domain of the forbidden coalition is annihilated. For mathematical convenience, we use a Karnaugh map as a sample space, in which the domain of the forbidden coalition is restored, albeit with zero content. Section 3 shows that, for a 2-out-of-3 system, the vector of total Banzhaf powers changes from a value of $\boldsymbol{TBP} = [2 \quad 2 \quad 2]^T$ to a value of $\boldsymbol{TBP} = [1 \quad 1 \quad 2]^T$ when the first two voters refuse to form a potentially valid coalition. Section 4 extends the analysis of Section 3 to an arbitrary k-out-of-n system, which serves as a general model for simple majority systems or super-majority systems, e.g., two-third, three-quarter, or four-fifth ones. We exemplify the arbitrary k-out-of-n system by a 5-out-of-8 one, visualized on an elegant eight-variable Karnaugh map. Our results demonstrate that the Public Good Index (PGI) is exactly equal to the Total Banzhaf Power (TBP) for a general k-out-of-n system, not only when this system is unrestricted, but also when it is restricted through the lack of co-operation between two voting members. Section 5 presents a scalar-weighted five-member voting system that represents the Scottish Parliament of 2007. Here, the TBP and the PGI cease to be exactly the same, but they both indicate a loss of power for the two largest parties when they refuse to form a coalition, with the ironic rise of the third largest party to the status of the most powerful one. For the examples in Sections 3 to 5, we checked our results by repeated calculation via independent means or via exhaustive visualization of the entire sample space, but for the example in Section 6, we had also a chance to check our results versus published ones. In fact, we reproduce the results of Yakuba [33] for a seven-member voting system in the unrestricted and restricted cases. Moreover, our results are visualized on an elegant seven-variable Karnaugh map. Section 7 discusses the generalization of the subject of exploration from one of a restriction on the formation of a specific coalition between exactly two voters to one of several restrictions on the formation of several coalitions that involve several voters each. Section 8 concludes the paper. To make the paper self-contained, its main text is supplemented with two appendices. Appendix A is an extensive introduction to the concept of a Boolean quotient, its general properties, its relations to the Boole-Shannon expansion and the Boolean difference, and its utility in interpreting various concepts of voting theory, especially those of voter desirability. Appendix B introduces and outlines prominent properties of k-out-of-n switching functions.

## 2. Switching-Algebraic Computation of Banzhaf and Banzhaf-Related Indices for General and Monotone Systems

To assess the power/influence of vote $X_m$ $(1 \leq m \leq n)$, the total Banzhaf power enumerates the number of states of the specific condition of resolution passing $f(\boldsymbol{X})$ subject to a supporting vote $X_m$ that swings to resolution rejection $\overline{f}(\boldsymbol{X})$ subject to a disapproving vote $\overline{X}_m$. Mathematically, this condition can be stated as a conjunction of two equations, *viz.*

$$\{f(\boldsymbol{X})/X_m = 1\} \cap \{\overline{f}(\boldsymbol{X})\,/\overline{X}_m = 1\}. \tag{1}$$

According to the Principle of Assertions [44-46], the above equational presentation is exactly equivalent to the propositional form or the switching function comprising the ANDing of two propositions:

$$(f(\boldsymbol{X})/X_m) \wedge (\overline{f}(\boldsymbol{X})\,/\overline{X}_m). \tag{2}$$

The expression in (2) is an indicator variable for the event that the vote $X_m$ is pivotal or critical in (correctly) determining the outcome $f(\boldsymbol{X})$. Here, the ratios $(f(\boldsymbol{X})/X_m)$ and $(\overline{f}(\boldsymbol{X})/\overline{X}_m)$ denote the Boolean quotients [44, 47-52] of the system function $f(\boldsymbol{X})$ w.r.t. the literal $X_m$ and of the complementary function $\overline{f}(\boldsymbol{X})$ w.r.t. the complementary literal $\overline{X}_m$, respectively, i.e.

$$f(\boldsymbol{X})/X_m = f(\boldsymbol{X}|(X_m = 1). \tag{3}$$

$$\overline{f}(\boldsymbol{X})/\overline{X}_m = \overline{f}(\boldsymbol{X}|(X_m = 0). \tag{4}$$

More details about Boolean quotients are given in Appendix A. Now, to enumerate the number of states of the specific propositional condition (2), we employ its weight as the total Banzhaf power $TBP(X_m)$ of voter number $m$, where we use the symbol $wt(h)$ to denote the weight or number of true vectors of a switching function $h$.

$$TBP(X_m) = wt((f(\boldsymbol{X})/X_m) \wedge (\overline{f}(\boldsymbol{X})\,/\overline{X}_m)), \qquad (1 \leq m \leq n). \tag{5}$$

Implicit in the definition above of the Banzhaf index (and in the definition of other indices in the sequel) is the assumption that system states or configurations are equally likely. This assumption in turn necessitates that voters cast their votes independently of each other. Each of the Boolean quotients $(f(\boldsymbol{X})/X_m)$ and $(\overline{f}(\boldsymbol{X})\,/\overline{X}_m)$, as well as their conjunction, is a function of $(n-1)$ variables, and hence, the weight of each of these three quantities belongs to the interval $[0,\ 2^{n-1}]$. We can divide $TBP(X_m)$ by $2^{n-1}$ to obtain a probabilistic Banzhaf power $PBP(X_m) = TBP(X_m)/2^{n-1}$ , which belongs to the unit interval $[0.0,\ 1.0]$, and that can be interpreted as the probability that the decision of voter $m$ is pivotal in determining the voting outcome $f(\boldsymbol{X})$. Formula (5) involves the weight of a specific

Boolean function $((f(\boldsymbol{X})/X_m) \wedge (\bar{f}(\boldsymbol{X})\,/\bar{X}_m))$, which is $2^{n-1}$ minus the weight of its complement $((\bar{f}(\boldsymbol{X})/X_m) \vee (\,f(\boldsymbol{X})/\bar{X}_m))$, that can be obtained from it via De Morgan's Law. Hence, $TBP(X_m)$ can also be expressed as

$$TBP(X_m) = 2^{n-1} - wt((\bar{f}(\boldsymbol{X})/X_m) \vee (\,f(\boldsymbol{X})/\bar{X}_m)), \qquad (1 \leq m \leq n). \tag{6a}$$

$$TBP(X_m) = 2^{n-1} - wt(\bar{f}(\boldsymbol{X})/X_m) - wt(\,f(\boldsymbol{X})/\bar{X}_m) + wt((\bar{f}(\boldsymbol{X})/X_m) \wedge (\,f(\boldsymbol{X})/\bar{X}_m)\,) = wt(f(\boldsymbol{X})/X_m) - wt(f(\boldsymbol{X})/\bar{X}_m) + wt((\bar{f}(\boldsymbol{X})/X_m) \wedge (\,f(\boldsymbol{X})/\bar{X}_m)\,) = wt\big(\bar{f}(\boldsymbol{X})/\bar{X}_m\big) - wt(\bar{f}(\boldsymbol{X})/X_m) + wt((\bar{f}(\boldsymbol{X})/X_m) \wedge (\,f(\boldsymbol{X})/\bar{X}_m)\,), \quad (1 \leq m \leq n). \tag{6b}$$

To obtain (6b) from (6a), we invoked the inclusion-exclusion principle [53-56] for two terms. Even though formulas (5) and (6) are a natural mathematical translation of the verbal definition of the Banzhaf index, they are almost unheard of in the open literature. These formulas (with their undesirable extra complication of complementing $f(\boldsymbol{X})$) are usually avoided since most voting systems are typically monotone. For a monotone voting system, the system function is a positively polarized unate function, and hence it is a monotonically non-decreasing one (monotonically increasing one, for short), i.e.,

$$(f(\boldsymbol{X})/X_m) \geq (f(\boldsymbol{X})/\bar{X}_m), \tag{7a}$$

or equivalently

$$(f(\boldsymbol{X})/\bar{X}_m) \wedge (\bar{f}(\boldsymbol{X})\,/X_m) = 0. \tag{7b}$$

Hence, formula (7b) might be ORed to formula (2) without changing its value, namely

$$(f(\boldsymbol{X})/X_m) \wedge (\bar{f}(\boldsymbol{X})\,/\bar{X}_m) = \big((f(\boldsymbol{X})/X_m) \wedge (\bar{f}(\boldsymbol{X})\,/\bar{X}_m)\big) \vee 0 = ((f(\boldsymbol{X})/X_m) \wedge (\bar{f}(\boldsymbol{X})\,/\bar{X}_m)) \vee ((f(\boldsymbol{X})/\bar{X}_m) \wedge (\bar{f}(\boldsymbol{X})\,/X_m)) = (f(\boldsymbol{X})/X_m) \oplus (f(\boldsymbol{X})/\bar{X}_m) = \frac{\partial f(\boldsymbol{X})}{\partial X_m} = \frac{\partial f(\boldsymbol{X})}{\partial \bar{X}_m} = \frac{\partial \bar{f}(\boldsymbol{X})}{\partial X_m} = \frac{\partial \bar{f}(\boldsymbol{X})}{\partial \bar{X}_m}. \tag{8}$$

Here, the symbol $\frac{\partial f(\boldsymbol{X})}{\partial X_m}$ denotes the partial derivative of the voting system Boolean function $f(\boldsymbol{X})$ w.r.t. its argument $X_m$ [57, 58]. Hence, total Banzhaf power in (5) can be rewritten in the following celebrated form, which is well-known for monotone voting systems [12, 15, 19, 21, 28, 29]

$$TBP(X_m) = wt\left(\frac{\partial f(\boldsymbol{X})}{\partial X_m}\right) = wt\left(\frac{\partial f(\boldsymbol{X})}{\partial \bar{X}_m}\right) = wt\left(\frac{\partial \bar{f}(\boldsymbol{X})}{\partial X_m}\right) = wt\left(\frac{\partial \bar{f}(\boldsymbol{X})}{\partial \bar{X}_m}\right). \qquad (1 \leq m \leq n). \tag{9}$$

For a monotone system, the fact that the function $f(\boldsymbol{X})$ is a positively polarized unate function can be further utilized to obtain more efficient formulas of $TBP(X_m)$ for $(1 \leq m \leq n)$ [12], namely

$$TBP(X_m) = 2\, wt(f(\boldsymbol{X})/X_m) - wt\big(f(\boldsymbol{X})\big). \tag{10}$$

$$TBP(X_m) = wt\big(f(\boldsymbol{X})\big) - 2\, wt(f(\boldsymbol{X})/\overline{X}_m). \tag{11}$$

$$TBP(X_m) = wt(f(\boldsymbol{X})/X_m) - wt(f(\boldsymbol{X})/\overline{X}_m). \tag{12}$$

$$TBP(X_m) = wt\big(\overline{f}(\boldsymbol{X})\big) - 2\, wt(\overline{f}(\boldsymbol{X})/X_m). \tag{13}$$

$$TBP(X_m) = 2\, wt(\overline{f}(\boldsymbol{X})/\overline{X}_m) - wt\big(\overline{f}(\boldsymbol{X})\big). \tag{14}$$

$$TBP(X_m) = wt\big(\overline{f}(\boldsymbol{X})/\overline{X}_m\big) - wt(\overline{f}(\boldsymbol{X})/X_m). \tag{15}$$

Note that if $f(\boldsymbol{X})$ is monotonically non-decreasing in its argument $X_m$, then $f(\boldsymbol{X}) = X_m\, f_{m1} \vee\ f_{m2}$, where the two functions $f_{m1} = f_{m1}(\boldsymbol{X}/X_m)$ and $f_{m2} = f_{m2}(\boldsymbol{X}/X_m)$ are functions of $\boldsymbol{X}$ excluding $X_m$. Hence, $f(\boldsymbol{X})/\overline{X}_m = f_{m2}$, $\overline{f}(\boldsymbol{X}) = (\overline{X}_m \vee \bar{f}_{m1})\bar{f}_{m2}$, and $\overline{f}(\boldsymbol{X})/X_m = \bar{f}_{m1}\bar{f}_{m2}$. This confirms the result $(\overline{f}(\boldsymbol{X})/X_m) \wedge (f(\boldsymbol{X})/\overline{X}_m) = 0$ in (7b), and hence (6b) leads (for a monotone system) to each of (12) and (15).

According to Appendix A, the Boole-Shannon expansions of $\overline{f}(\boldsymbol{X})$ and $f(\boldsymbol{X})$ are

$$\overline{f}(\boldsymbol{X}) = (\overline{f}(\boldsymbol{X})/X_m) \wedge X_m \vee (\overline{f}(\boldsymbol{X})\,/\overline{X}_m) \wedge \overline{X}_m, \tag{16}$$

$$f(\boldsymbol{X}) = (f(\boldsymbol{X})/X_m) \wedge X_m \vee (f(\boldsymbol{X})\,/\overline{X}_m) \wedge \overline{X}_m, \tag{17}$$

These expansions allow us to compute the weights $wt\big(\overline{f}(\boldsymbol{X})\big)$ and $wt\big(f(\boldsymbol{X})\big)$ of the two functions $\overline{f}(\boldsymbol{X})$ and $f(\boldsymbol{X})$, which add to $2^n$, as

$$wt\,\big(\overline{f}(\boldsymbol{X})\big) = wt\big(\overline{f}(\boldsymbol{X})/X_m\big) + wt\big(\overline{f}(\boldsymbol{X})/\,\overline{X}_m\big). \tag{18}$$

$$wt\,\big(f(\boldsymbol{X})\big) = wt(f(\boldsymbol{X})/X_m) + wt\big(f(\boldsymbol{X})/\,\overline{X}_m\big), \tag{19}$$

Equations (18) and (19) are included herein to serve the purpose of comparison with earlier formulas, and to verify consistency among (10)-(15). The best way to obtain the weight of a switching function $h(\boldsymbol{X})$ is to express this function as a sum of disjoint products [53, 54, 59-62] $h(\boldsymbol{X}) = \bigvee_{k=1}^{m} D_k$, where $D_i \wedge D_j = 0, \forall$ i, j, for then the weight of $h(\boldsymbol{X})$ is [63-65]

$$wt(h(\boldsymbol{X})) = \textstyle\sum_{k=1}^{m} wt(D_k) = \sum_{k=1}^{m} 2^{(n-\ell(D_k))}. \tag{20}$$

where the weight of a product $D_k$ is equal to $2^{(n-\ell(P_k))}$, and the symbol $\ell(P_k)$

denotes the number of irredundant literals in the product $D_k$, e.g. $\ell(1) = 0$, $\ell(X_i) = \ell(\overline{X}_i) = 1$, $\ell(X_i X_j) = \ell(X_i \overline{X}_j) = 2$, while $\ell(0)$ is assumed to be infinity. Casting a function in a disjoint s-o-p form not only facilitates the computation of its weight, but it also facilitates the computation of the weights of all its Boolean quotients, since the construction of a Boolean quotient preserves disjointness [12].

Two interesting variations of the Banzhaf index are the power-to-initiate index (PII), and the power-to-prevent index (PPI) [66]. The power-to-initiate index (PII) for an individual voter $m$ can be defined as the conditional probability that the voter's decision $X_m$ is pivotal in determining the voting outcome $f(\boldsymbol{X})$ given that a negative decision $\overline{f}(\boldsymbol{X})$ has been reached. Recalling that the indicator for voter's decision $X_m$ being pivotal is given by (2), while that of a negative decision is expressed by $\overline{f}(\boldsymbol{X})$, and noting that the weight of a switching function behaves exactly like a probability [61, 63-65], we write

$$PII(X_m) = wt((f(\boldsymbol{X})/X_m) \wedge (\overline{f}(\boldsymbol{X})/\overline{X}_m) \wedge \overline{f}(\boldsymbol{X}))/wt(\overline{f}(\boldsymbol{X})), \qquad (1 \leq m \leq n). \tag{21}$$

The power-to-prevent index (PPI) for an individual voter $m$ can be defined as the conditional probability that the voter's decision is pivotal in determining the voting outcome $f(\boldsymbol{X})$ given that a positive decision $f(\boldsymbol{X})$ has been reached, namely

$$PPI(X_m) = wt((f(\boldsymbol{X})/X_m) \wedge (\overline{f}(\boldsymbol{X})/\overline{X}_m) \wedge f(\boldsymbol{X}))/wt(f(\boldsymbol{X})), \qquad (1 \leq m \leq n). \tag{22}$$

Thanks to the Boole-Shannon expansions (16) and (17) and to the orthogonality of $(f(\boldsymbol{X})/X_m)$ and $(\overline{f}(\boldsymbol{X})/X_m)$, as well as that of $(f(\boldsymbol{X})/\overline{X}_m)$ and $(\overline{f}(\boldsymbol{X})/\overline{X}_m)$ (See Appendix A), formulas (19) and (20) reduce to

$$PII(X_m) = wt((f(\boldsymbol{X})/X_m) \wedge (\overline{f}(\boldsymbol{X})/\overline{X}_m) \wedge \overline{X}_m)/wt(\overline{f}(\boldsymbol{X})) = wt((f(\boldsymbol{X})/X_m) \wedge (\overline{f}(\boldsymbol{X})/\overline{X}_m))\, wt(\overline{X}_m)/wt(\overline{f}(\boldsymbol{X})) = wt((f(\boldsymbol{X})/X_m) \wedge (\overline{f}(\boldsymbol{X})/\overline{X}_m))/wt(\overline{f}(\boldsymbol{X})), \quad (1 \leq m \leq n). \tag{23}$$

$$PPI(X_m) = wt((f(\boldsymbol{X})/X_m) \wedge (\overline{f}(\boldsymbol{X})/\overline{X}_m) \wedge X_m)/wt(f(\boldsymbol{X})) = wt((f(\boldsymbol{X})/X_m) \wedge (\overline{f}(\boldsymbol{X})/\overline{X}_m))\, wt(X_m)/wt(f(\boldsymbol{X})) = wt((f(\boldsymbol{X})/X_m) \wedge (\overline{f}(\boldsymbol{X})/\overline{X}_m))/wt(f(\boldsymbol{X})), \quad (1 \leq m \leq n). \tag{24}$$

Note that each of the four Boolean quotients in (16) and (17) is independent of $X_m$ (thanks to independence among components of $\boldsymbol{X}$). Now, we obtain

$$(PII(X_m))^{-1} + (PPI(X_m))^{-1} = (wt(\overline{f}(\boldsymbol{X}) + wt(f(\boldsymbol{X}))/(wt((f(\boldsymbol{X})/X_m) \wedge (\overline{f}(\boldsymbol{X})/\overline{X}_m)) = 2^n/(wt((f(\boldsymbol{X})/X_m) \wedge (\overline{f}(\boldsymbol{X})/\overline{X}_m)) = 2\,(PBI(X_m))^{-1}, \quad (1 \leq m \leq n), \tag{25}$$

which means that the probabilistic Banzhaf index is the *harmonic mean* [67] of the power-to-initiate index, and the power-to-prevent index.

Another type of Banzhaf-related voting power measure is the satisfaction index $SAT(X_m)$ [68-72], which is the probability that voter $m$ is satisfied with the system decision, i.e, the probability that the vote $X_m$ coincides (or is equivalent to) the decision function $f(\boldsymbol{X})$, *viz.*

$$SAT(X_m) = wt(X_m \odot f(\boldsymbol{X}))/2^n = wt(X_m f(\boldsymbol{X}) \vee \overline{X}_m \overline{f}(\boldsymbol{X}))/2^n = (wt(X_m f(\boldsymbol{X})) + wt(\overline{X}_m \overline{f}(\boldsymbol{X})))/2^n = (wt(X_m(f(\boldsymbol{X})/X_m)) + wt(\overline{X}_m(\overline{f}(\boldsymbol{X})/\overline{X}_m)))/2^n = (wt(f(\boldsymbol{X})/X_m) + wt((\overline{f}(\boldsymbol{X})/\overline{X}_m))/2^n, \quad (1 \leq m \leq n), \tag{26}$$

where we made use of the expansions (17) and (18), the facts that the Boolean quotients $(f(\boldsymbol{X})/X_m)$ and $(\overline{f}(\boldsymbol{X})/\overline{X}_m)$ are independent of $X_m$ and $\overline{X}_m$ and the relations $wt(X_m) = wt(\overline{X}_m) = 1$. The function $(X_m \odot f(\boldsymbol{X}))$ is of $n$ variables, and its weight must be divided by $2^n$ to achieve a probability interpretation.

The negatively oriented satisfaction index $NSAT(X_m)$ for a voter $m$ is defined (in an analogous way to that of the PII) as the conditional probability that voter $m$ is satisfied given that the voting decision is negative [66], or in switching-algebraic terms:

$$NSAT(X_m) = wt((X_m \odot f(\boldsymbol{X})) \wedge \overline{f}(\boldsymbol{X}))/ wt(\overline{f}(\boldsymbol{X})) = wt(\overline{X}_m \overline{f}(\boldsymbol{X}))/wt(\overline{f}(\boldsymbol{X})) = wt(\overline{X}_m(\overline{f}(\boldsymbol{X})/\overline{X}_m))/wt(\overline{f}(\boldsymbol{X}) = wt((\overline{f}(\boldsymbol{X})/\overline{X}_m))/wt(\overline{f}(\boldsymbol{X})), \quad (1 \leq m \leq n). \tag{27}$$

The positively oriented satisfaction index $PSAT(X_m)$ for a voter $m$ is defined (in an analogous way to that of the PPI) as the conditional probability that voter $m$ is satisfied given that the voting decision is positive [66], namely

$$PSAT(X_m) = wt((X_m \odot f(\boldsymbol{X})) \wedge f(\boldsymbol{X}))/ wt(f(\boldsymbol{X})) = wt(X_m f(\boldsymbol{X}))/wt(f(\boldsymbol{X})) = (wt(X_m(f(\boldsymbol{X})/X_m))/wt(f(\boldsymbol{X}) = (wt(f(\boldsymbol{X})/X_m)/wt(f(\boldsymbol{X})), \quad (1 \leq m \leq n). \tag{28}$$

Since $wt(\overline{f}(\boldsymbol{X}))$ and $wt(f(\boldsymbol{X}))$ add to $2^n$, equations (26)-(28) indicate that $SAT(X_m)$ is a *weighted arithmetic mean* of $NSAT(X_m)$ and $PSAT(X_m)$ [73]. The satisfaction index $SAT(X_m)$ is also related to the probabilistic Banzhaf power, deduced from (12) in the monotone case, since

$$SAT(X_m) = (wt(f(\boldsymbol{X})/X_m) + wt((\overline{f}(\boldsymbol{X})/\overline{X}_m))/2^n = (wt(f(\boldsymbol{X})/X_m) + 2^{n-1} - wt((\overline{f}(\boldsymbol{X})/X_m))/2^n = (1 + PBP(X_m))/2, \quad (1 \leq m \leq n). \tag{29}$$

There are many other related power indices. For example, Napel and Widgrén [74] strengthened the concept of 'dummy voters' by introducing that of 'inferior players' and argued according to their novel paradigm that only non-inferior voters can be viewed as powerful. These authors proposed a new voting index called the Strict Power Index (SPI), such that $SPI(X_i) = TBP(X_i)$ for powerful members, and $SPI(X_i) = 0$ for inferior ones. Another interesting index is the Public Good Index (PGI), introduced by Holler [75] and axiomatized by Holler and Packel [76]. The PGI looks at $f(\boldsymbol{X})$ so as to count the number of the products representing the $MWCs$ in which the uncomplemented literal of a certain voter appears and take this number as the power of this voter [12, 77]. The PGI is perhaps the simplest existing index, and is definitely useful for the topic of the present paper when restrictions are imposed on coalition formation. It seems that it has not gained the popularity it deserves, because it did not meet sophistication standards unfortunately demanded by some meticulous scholars, and it did not conform to certain unjustified suppositions based solely on intuition. Our forthcoming examples obviously contradict the (unfortunately popular) opinion that the PGI is too simple to be remarkably useful.

## 3. A 2-out-of-3 Voting System

Many scalar-weighted voting systems (of a wide variety of weights and quota), such as the systems [2; 1, 1, 1], [50; 33, 33, 33], [51; 48, 47, 6] and [50; 49, 49, 1] are equivalent to the 2-out-of-3 system (also called the triple modular redundancy), specified by either its (self-dual) decision function $f(\boldsymbol{X})$ or its complement $\overline{f}(\boldsymbol{X})$, namely

$$f(\boldsymbol{X}) = X_1X_2 \ \vee \ X_2X_3 \ \vee \ X_1X_3, \tag{30}$$

$$\overline{f}(\boldsymbol{X}) = \overline{X}_1\overline{X}_2 \ \vee \overline{X}_2\overline{X}_3 \ \vee \overline{X}_1\overline{X}_3. \tag{31}$$

Due to total symmetry within either function (and despite frequent absence of such symmetry within the weights), the Banzhaf indices must be the same, i.e., $TBP(X_1) = TBP(X_2) = TBP(X_3)$. Hence, it suffices to compute the Boolean quotient or derivative w.r.t. one of the variables $X_1, X_2$ and $X_3$ (say $X_1$). To facilitate further processing, we rewrite each of the decision function and its complement in the form of a sum of disjoint products, and then replace the OR operator (∨) by an XOR operator (⊕) in each of these sums. Mathematically, we obtain:

$$f(\boldsymbol{X}) = X_1X_2 \vee \overline{X}_1X_2X_3 \vee X_1\overline{X}_2X_3 = X_1X_2 \oplus \overline{X}_1X_2X_3 \oplus X_1\overline{X}_2X_3. \tag{32}$$

$$\overline{f}(\boldsymbol{X}) = \overline{X}_1\overline{X}_2 \ \vee X_1\overline{X}_2\overline{X}_3 \ \vee \overline{X}_1X_2\overline{X}_3 = \overline{X}_1\overline{X}_2 \ \oplus X_1\overline{X}_2\overline{X}_3 \ \oplus \overline{X}_1X_2\overline{X}_3. \tag{33}$$

Noting that $wt(f) = wt(\overline{f}) = 4$, we compute the Banzhaf index $TBP(X_1)$ repeatedly via (5), (6) and (9)-(15), as a way of demonstrating and verifying the plethora of formulas we have:

$$TBP(X_1) = wt((f/X_1)(\overline{f}/\overline{X}_1)) = wt((X_2 \oplus \overline{X}_2X_3)(\overline{X}_2 \ \oplus X_2\overline{X}_3)) = wt(\overline{X}_2X_3 \ \oplus X_2\overline{X}_3) = 1 + 1 = 2. \tag{34}$$

$$TBP(X_1) = 2^{n-1} - wt((\overline{f}/X_1) \vee (f/\overline{X}_1)) = 4 - wt(\overline{X}_2\overline{X}_3 \vee X_2X_3) = 4 - 2 = 2. \quad (35)$$

$$TBP(X_1) = wt(\frac{\partial f}{\partial X_1}) = wt(X_2 \oplus X_2X_3 \oplus \overline{X}_2X_3) = wt(X_2\overline{X}_3 \oplus \overline{X}_2X_3) = 2. \quad (36)$$

$$TBP(X_1) = 2\, wt(f/X_1) - wt(f) = 2\, wt\big(X_2 \oplus \overline{X}_2X_3\big) - 4 = 2(2+1) - 4 = 2. \quad (37)$$

$$TBP(X_1) = wt(f) - 2\, wt(f/\overline{X}_1) = 4 - 2\, wt(X_2X_3) = 4 - 2(1) = 2. \quad (38)$$

$$TBP(X_1) = wt(f/X_1) - wt\big(f/\overline{X}_1\big) = wt\big(X_2 \oplus \overline{X}_2X_3\big) - wt(X_2X_3) = (2+1) - 1 = 2. \quad (39)$$

$$TBP(X_1) = wt\big(\overline{f}\big) - 2\, wt\big(\overline{f}/X_1\big) = 4 - 2\, wt\big(\overline{X}_2\overline{X}_3\big) = 4 - 2(1) = 2. \quad (40)$$

$$TBP(X_1) = 2\, wt\big(\overline{f}/\overline{X}_1\big) - wt\big(\overline{f}\big) = 2\, wt\big(\overline{X}_2 \oplus X_2\overline{X}_3\big) - 4 = 2(2+1) - 4 = 2. \quad (41)$$

$$TBP(X_1) = wt\big(\overline{f}/\overline{X}_1\big) - wt\big(\overline{f}/X_1\big) = wt\big(\overline{X}_2 \oplus X_2\overline{X}_3\big) - wt\big(\overline{X}_2\overline{X}_3\big) = (2+1) - 1 = 2. \quad (42)$$

Now, we consider the case where the coalition $X_1X_2$ is forbidden. We need to impose the constraint $\{X_1X_2 = 0\}$ on the function $f(\boldsymbol{X})$, which leads to replacing $f(\boldsymbol{X})$ by another function, say $g(\boldsymbol{X})$. First, we construct the Boole-Shannon expansion of $f(\boldsymbol{X})$ w.r.t. the two variables $X_1$ and $X_2$, namely

$$f(\boldsymbol{X}) = \big(f(\boldsymbol{X})/\overline{X}_1\overline{X}_2\big)\, \overline{X}_1\overline{X}_2 \oplus \big(f(\boldsymbol{X})/\overline{X}_1X_2\big)\, \overline{X}_1X_2 \oplus \big(f(\boldsymbol{X})/X_1\overline{X}_2\big)\, X_1\overline{X}_2 \oplus (f(\boldsymbol{X})/X_1X_2)\, X_1X_2. \quad (43)$$

$$f(\boldsymbol{X}) = (0)\, \overline{X}_1\overline{X}_2 \oplus (X_3)\, \overline{X}_1X_2 \oplus (X_3)\, X_1\overline{X}_2 \oplus (1)\, X_1X_2. \quad (44)$$

Now we construct $g(\boldsymbol{X})$ so as to share all subfunctions with $f(\boldsymbol{X})$, with the exception of the subfunction w.r.t. $X_1X_2$, which is nullified.

$$g(\boldsymbol{X}) = (0)\, \overline{X}_1\overline{X}_2 \oplus (X_3)\, \overline{X}_1X_2 \oplus (X_3)\, X_1\overline{X}_2 \oplus (\boldsymbol{0})\, X_1X_2 = \overline{X}_1X_2X_3 \oplus X_1\overline{X}_2X_3. \quad (45)$$

We obtain the complementary function $\overline{g}(\boldsymbol{X})$ by complementing each subfunction of $g(\boldsymbol{X})$, i.e. complementing each subfunction of $f(\boldsymbol{X})$ with the exception of the subfunction w.r.t. $X_1X_2$, which is now asserted.

$$\overline{g}(\boldsymbol{X}) = (1)\, \overline{X}_1\overline{X}_2 \oplus \big(\overline{X}_3\big)\, \overline{X}_1X_2 \oplus \big(\overline{X}_3\big)\, X_1\overline{X}_2 \oplus (\boldsymbol{1})\, X_1X_2 = \overline{X}_1\overline{X}_2 \oplus \overline{X}_1X_2\overline{X}_3 \oplus X_1\overline{X}_2\overline{X}_3 \oplus X_1X_2. \quad (46)$$

Note that the functions $g(\boldsymbol{X})$ and $\overline{g}(\boldsymbol{X})$ satisfy:

$$g(\boldsymbol{X}) \wedge \overline{g}(\boldsymbol{X}) = 0, \quad (47)$$

$$g(\boldsymbol{X}) \vee \overline{g}(\boldsymbol{X}) = 1, \quad (48)$$

as required for any complementary functions, albeit being implicitly subject to the constraint $X_1X_2 = 0$. Note that the product $X_1X_2$ actually does not exist.

Note also that the functions $g(\boldsymbol{X})$ and $\overline{g}(\boldsymbol{X})$ are no longer unate; each of them is biform in $X_1$ and $X_2$, while remaining monoform in $X_3$. The functions $g(\boldsymbol{X})$ and $\overline{g}(\boldsymbol{X})$ lost the total symmetry of $f(\boldsymbol{X})$ and $\overline{f}(\boldsymbol{X})$, but each of them remains partially symmetric in $X_1$ and $X_2$, and hence $TBP(X_1) = TBP(X_2)$, and it suffices to compute one of these two values, say $TBP(X_1)$. Since $g(\boldsymbol{X})$ is no longer monotone in $X_1$, formulas (5) and (6) are the only valid formulas for computing $TBP(X_1)$, which turn out, respectively, to be

$$TBP(X_1) = wt((g/X_1)(\overline{g}/\overline{X}_1)) = wt((\overline{X}_2X_3)(\overline{X}_2 \oplus X_2\overline{X}_3)) = wt(\overline{X}_2X_3) = 1. \quad (49)$$

$$TBP(X_1) = 2^{n-1} - wt((\overline{g}/X_1) \vee (g/\overline{X}_1)) = 4 - wt((\overline{X}_2\overline{X}_3)\vee(X_2X_3)) = 4 - wt(\overline{X}_2\overline{X}_3\vee X_2\vee X_2X_3) = 4 - wt(\overline{X}_2\overline{X}_3\vee X_2) = 4 - wt(\overline{X}_2\overline{X}_3 \oplus X_2) = 4 - (1+2) = 1. \quad (50)$$

By contrast, $TBP(X_3)$ might be computed via any of the formulas (5), (6) and (9)-(15). For example, formula (5) yields

$$TBP(X_3) = wt((g/X_3)(\overline{g}/\overline{X}_3)) = wt((\overline{X}_1X_2 \oplus X_1\overline{X}_2)(\overline{X}_1\overline{X}_2 \oplus \overline{X}_1X_2 \oplus X_1\overline{X}_2 \oplus X_1X_2)) = wt((\overline{X}_1X_2 \oplus X_1\overline{X}_2)) = 2. \quad (51)$$

Alternatively, formula (12) that solely uses $g(\boldsymbol{X})$ obtains

$$TBP(X_3) = wt(g/X_3) - wt(g/\overline{X}_3) = wt((\overline{X}_1X_2 \oplus X_1\overline{X}_2)) - wt(0) = 2 - 0 = 2. \quad (52)$$

Likewise, formula (15) that solely uses $\overline{g}(\boldsymbol{X})$ obtains:

$$TBP(X_3) = wt(\overline{g}/\overline{X}_3) - wt(\overline{g}/X_3) = wt(\overline{X}_1\overline{X}_2 \oplus \overline{X}_1X_2 \oplus X_1\overline{X}_2 \oplus X_1X_2) - wt(\overline{X}_1\overline{X}_2 \oplus X_1X_2) = 4 - 2 = 2, \quad (53)$$

Note that if we insist that $X_1X_2 = 0$ in (46) (or if we have assigned the value 0 instead of 1 to the subfunction of $\overline{g}(\boldsymbol{X})$ w.r.t. $X_1X_2$,), then we still get $TBP(X_3) = 3 - 1 = 2$. A visual demonstration of the above results can be obtained from the probability sample spaces of Figs. 1 and 2. Figure 1 is the true sample space, while Fig. 2 is an adequate one that we are employing for the sake of mathematical simplicity, and it works provided we adhere to the use of permissible formulas only for voting powers. We note that the vector of total Banzhaf powers changed from a value of $\boldsymbol{TBP} = [2 \quad 2 \quad 2]^T$ to a value of $\boldsymbol{TBP} = [1 \quad 1 \quad 2]^T$ when the first two voters refused to form the potentially valid coalition including them. This result agrees with intuition; those who refuse to cooperate are diminishing their powers. It is ironic to view the [50; 49, 49, 1] system. Normally, the 1-seat party is as powerful as any of the 49-seat parties, but (as if the foregoing surprise is not enough). When these two large parties refuse to talk to and cooperate with each other, the much smaller party becomes twice as powerful as any of the two much larger ones.

In passing, we note that we can (by just glancing at (30) deduce that the Public Good Index is normally of a value $\boldsymbol{PGI} = [2 \quad 2 \quad 2]^T$, and that this value changes to $\boldsymbol{PGI} = [1 \quad 1 \quad 2]^T$ upon omitting the prime implicant $X_1X_2$ in (30). It might seem surprising that the numerical values of the $\mathrm{PGI}$ and the $\mathrm{TBP}$ are in exact agreement in both the normal and restricted cases. In the sequel, we show that $PGI(X_m)$ is exactly equal to $TBP(X_m)$ for a general k-out-of-n system, whether this system is unrestricted, or it is restricted through the lack of co-operation between two voting members. We observe that it seems likely that $PGI(X_m)$ is indeed equivalent to $TBP(X_m)$ for a general k-out-of-n system under any type of restriction, but currently this is only a conjecture, and it remains to find a formal justification or proof for it.

| | $\overline{X}_3$ | $X_3$ | |
|---|---|---|---|
| $\overline{X}_1$ | 0 | 0 | $\overline{X}_2$ |
| $\overline{X}_1$ | 0 | 1 | $X_2$ |
| $X_1$ | 0 | 1 | $\overline{X}_2$ |

$g(\boldsymbol{X})$

Figure 1. The exact Universe of Discourse (probability sample space) that describes the new decision function $g(\boldsymbol{X})$ when the coalition $X_1X_2$ is not allowed in Example 1. Here, the domain $X_1X_2 = 1$ is annihilated, and statistical independence between $X_1$ and $X_2$ is lost. The sample space is a Karnaugh-map like-structure, but it is not a Karnaugh map *per se*. The function $g(\boldsymbol{X})$ ceases to be monotonically non-decreasing in each of $X_1$ and $X_2$ but continues to be so for $X_3$. With the coalition $X_1X_2$ being forbidden, total symmetry is spoiled, but partial symmetry between $X_1$ and $X_2$ is still retained. Furthermore, $TBP(X_1) = 1$ since there is a single solution ($X_2 = 0, X_3 = 1$) for the equation $g(1, X_2, X_3)\, \overline{g}(0, X_2, X_3) = 1$ (a single swing from a blue cell to an adjacent red one across the $X_1$ boundary), while $TBP(X_3) = 2$ since there are two solutions ($X_1 = 0, X_2 = 1$) and ($X_1 = 1, X_2 = 0$), both guaranteed to give $X_3 = 1$, for the equation $g(X_1, X_2, X_3)\, \overline{g}\big(X_1, X_2, \overline{X_3}\big) = 1$.

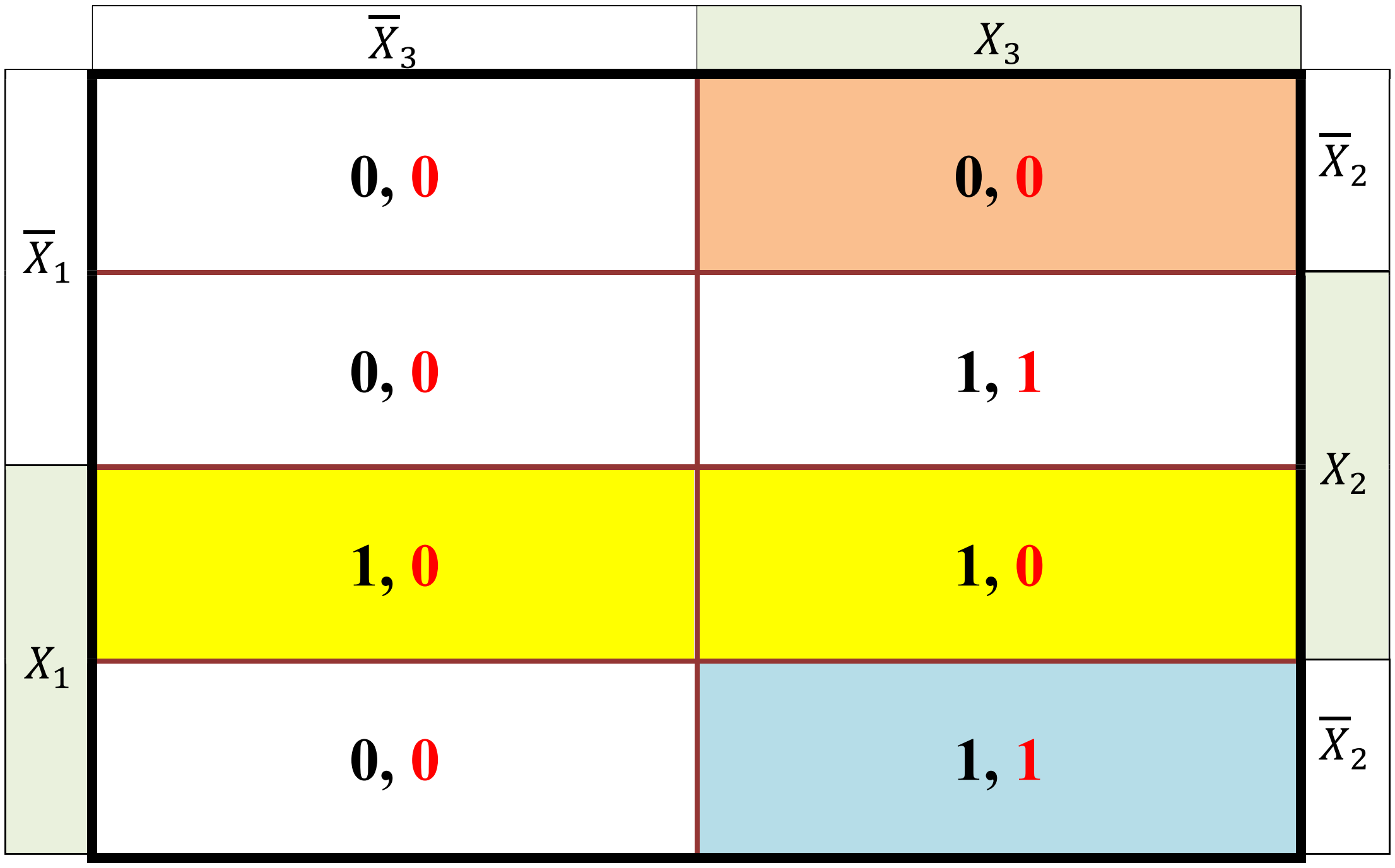

$f(\boldsymbol{X}), g(X)$

Figure 2. The Karnaugh map of the two decision functions in Example 1. The original decision function $f(\boldsymbol{X})$ is a monotonically non-decreasing one. When a coalition $X_1X_2$ is not allowed, it is mathematically convenient to obtain the new decision function $g(\boldsymbol{X})$ from the original one by forcing the two yellow cells (of loop $X_1X_2$) to be 0's for $g(\boldsymbol{X})$. Again, the function $g(\boldsymbol{X})$ ceases to be monotonically non-decreasing in each of $X_1$ and $X_2$ but continues to be so for $X_3$. Again, with the coalition $X_1X_2$ being forbidden, total symmetry is spoiled. Furthermore, $TBP(X_1) = 1$ since there is a single solution $(X_2 = 0, X_3 = 1)$ for the equation $g(1, X_2, X_3)\,\overline{g}(0, X_2, X_3) = 1$ (a single swing from a blue cell to an adjacent red one across the $X_1$ boundary), while $TBP(X_3) = 2$ since there are two solutions $(X_1 = 0, X_2 = 1)$ and $(X_1 = 1,\ X_2 = 0)$, both guaranteed to have $X_3 = 1$, for the equation $g(X_1, X_2, X_3)\,\overline{g}(X_1, X_2, \overline{X_3}) = 1$. The reverse swing from the arbitrarily assigned value of $g(1, 1, 1) = 0$ to the true value $g(0, 1, 1) = 1$ does not contribute to $TBP(X_1)$. This is guaranteed by using (6a) or (6b) (but none of (9)-(15)) to compute $TBP(X_1)$.

## 4. A General k-out-of-n Voting System

In this section, we generalize the analysis of the previous section by considering an arbitrary k-out-of-n system ($n \geq 3, k \geq 2$). From Appendix B, we know that the decision function $f(\boldsymbol{X})$ for this system is the monotonically non-decreasing symmetric switching function $Sy(n;\{k..n\};\boldsymbol{X})$, and that $TBP(X_m) = PGI(X_m) = c(n-1, k-1)$ for each voter $X_m$ in the k-out-of-n voting system. Now, we consider the case where no coalition is allowed to include $X_1$ and $X_2$ together. Unless $k = 2$, the coalition $X_1X_2$ is not a minimal winning coalition, but it is part of several such coalitions. The restriction that no coalition be allowed to include $X_1$ and $X_2$ together spoils the original total symmetry of $f(\boldsymbol{X})$. However, partial symmetry is retained between $X_1$ and $X_2$ and also among all the remaining voters. Therefore, $TBP(X_1) = TBP(X_2)$, while $TBP(X_m)$ is the same for $m = 3, 4, \ldots, n$. Hence, it suffices to compute the voting power for one of the two variables $X_1$ and $X_2$ (say $X_1$), and one of the remaining variables (say $X_3$).

Again, we need to impose the constraint $\{X_1X_2 = 0\}$ on the decision function $f(\boldsymbol{X})$, which leads to replacing $f(\boldsymbol{X})$ by another function, say $g(\boldsymbol{X})$. First, we construct the Boole-Shannon expansion of $f(\boldsymbol{X})$ w.r.t. the two variables $X_1$ and $X_2$, namely

$$f(\boldsymbol{X}) = \left(f(\boldsymbol{X})/\overline{X}_1\overline{X}_2\right)\overline{X}_1\overline{X}_2 \oplus \left(f(\boldsymbol{X})/\overline{X}_1X_2\right)\overline{X}_1X_2 \oplus \left(f(\boldsymbol{X})/X_1\overline{X}_2\right)X_1\overline{X}_2 \oplus \left(f(\boldsymbol{X})/X_1X_2\right)X_1X_2. \qquad (54)$$

Here, the four Boolean quotients in (54) are given (for $\boldsymbol{X_r} = \boldsymbol{X}/X_1X_2$ ) by

$$f(\boldsymbol{X})/\overline{X}_1\overline{X}_2 = Sy(n;\{k..n\};\boldsymbol{X})/\overline{X}_1\overline{X}_2 = Sy(n-2;\{k..(n-2)\};\boldsymbol{X_r})$$

$$f(\boldsymbol{X})/\overline{X}_1X_2 = Sy(n;\{k..n\};\boldsymbol{X})/\overline{X}_1X_2 = Sy(n-2;\{(k-1)..(n-2)\};\boldsymbol{X_r})$$

$$f(\boldsymbol{X})/X_1\overline{X}_2 = Sy(n;\{k..n\};\boldsymbol{X})/X_1\overline{X}_2 = Sy(n-2;\{(k-1)..(n-2)\};\boldsymbol{X_r})$$

$$f(\boldsymbol{X})/X_1X_2 = Sy(n;\{k..n\};\boldsymbol{X})/X_1X_2 = Sy(n-2;\{(k-2)..(n-2)\};\boldsymbol{X_r})$$

Hence, the function $f(\boldsymbol{X})$ is given by

$$f(\boldsymbol{X}) = \left(Sy(n-2;\{k..(n-2)\};\boldsymbol{X_r})\right)\overline{X}_1\overline{X}_2 \oplus \left(Sy(n-2;\{(k-1)..(n-2)\};\boldsymbol{X_r})\right)\overline{X}_1X_2 \oplus \left(Sy(n-2;\{(k-1)..(n-2)\};\boldsymbol{X_r})\right)X_1\overline{X}_2 \oplus \left(Sy(n-2;\{(k-2)..(n-2)\};\boldsymbol{X_r})\right)X_1X_2. \qquad (55)$$

Now we construct $g(\boldsymbol{X})$ so as to share all subfunctions with $f(\boldsymbol{X})$, with the exception of the subfunction w.r.t. $X_1X_2$, which is nullified

$$g(\boldsymbol{X}) = \left(Sy(n-2;\{k..(n-2)\};\boldsymbol{X_r})\right)\overline{X}_1\overline{X}_2 \oplus \left(Sy(n-2;\{(k-1)..(n-2)\};\boldsymbol{X_r})\right)\overline{X}_1X_2 \oplus \left(Sy(n-2;\{(k-1)..(n-2)\};\boldsymbol{X_r})\right)X_1\overline{X}_2 \oplus (\boldsymbol{0})\,X_1X_2. \qquad (56)$$

$$g(\boldsymbol{X}) = \big(Sy(n-2;\{k..(n-2)\};\boldsymbol{X_r})\big)\,\overline{X}_1\overline{X}_2 \oplus \big(Sy(n-2;\{(k-1)..(n-2)\};\boldsymbol{X_r})\big)\,(\overline{X}_1X_2 \oplus X_1\overline{X}_2). \quad (57)$$

Noting that the complement of a symmetric switching function (SSF) is another SSF whose characteristic set is the complementary characteristic set w.r.t. $\boldsymbol{I}_{n+1} = \{0, 1, 2, \ldots, n\}$, we obtain the complementary function $\overline{g}(\boldsymbol{X})$ by complementing each subfunction of $g(\boldsymbol{X})$, or complementing each subfunction of $f(\boldsymbol{X})$ with the exception of the subfunction w.r.t. $X_1X_2$, which is now asserted.

$$\overline{g}(\boldsymbol{X}) = \big(Sy(n-2;\{0..(k-1)\};\boldsymbol{X_r})\big)\,\overline{X}_1\overline{X}_2 \oplus \big(Sy(n-2;\{0..(k-2)\};\boldsymbol{X_r})\big)\,(\overline{X}_1X_2 \oplus X_1\overline{X}_2) \oplus (\boldsymbol{1})\,X_1X_2. \quad (58)$$

Note also that the functions $g(\boldsymbol{X})$ and $\overline{g}(\boldsymbol{X})$ are no longer unate; each of them is biform in $X_1$ and $X_2$, but each of them remains monoform in each of the remaining variables. Since $g(\boldsymbol{X})$ is no longer monotone in $X_1$, formulas (5) and (6) are the only valid formulas for computing $TBP(X_1) = TBP(X_2)$, which turn out, respectively, to be

$$TBP(X_1) = wt((g/X_1)(\overline{g}/\overline{X}_1)) = wt(Sy(n-2;\{(k-1)..(n-2)\};\boldsymbol{X_r})\overline{X}_2\,(Sy(n-2;\{0..(k-1)\};\boldsymbol{X_r})\,\overline{X}_2 \oplus Sy(n-2;\{0..(k-2)\};\boldsymbol{X_r})\,X_2)) = wt(Sy(n-2;\{(k-1)\};\boldsymbol{X_r})\overline{X}_2) = c(n-2,k-1). \quad (59)$$

$$TBP(X_1) = 2^{n-1} - wt((\overline{g}/X_1) \vee (\,g/\overline{X}_1)) = 2^{n-1} - wt((Sy(n-2;\{0..(k-2)\};\boldsymbol{X_r})\overline{X}_2 \oplus X_2) \vee (\big(Sy(n-2;\{k..(n-2)\};\boldsymbol{X_r})\big)\,\overline{X}_2 \oplus \big(Sy(n-2;\{(k-1)..(n-2)\};\boldsymbol{X_r})\big)\,X_2\,)). \quad (60)$$

$$TBP(X_1) = 2^{n-1} - wt\big(Sy(n-2;\{0..(k-2)\};\boldsymbol{X_r})\overline{X}_2\big) - 2^{n-2} - wt\big(Sy(n-2;\{k..(n-2)\};\boldsymbol{X_r})\,\overline{X}_2\big) - wt(Sy(n-2;\{(k-1)..(n-2)\};\boldsymbol{X_r})X_2\,) + wt(Sy(n-2;\{(k-1)..(n-2)\};\boldsymbol{X_r})X_2\,) = 2^{n-1} - (2^{n-2} - wt\big(Sy(n-2;\{(k-1)\};\boldsymbol{X_r})\big) - 2^{n-2} = wt\big(Sy(n-2;\{(k-1)\};\boldsymbol{X_r})\big) = c(n-2,k-1). \quad (61)$$

By contrast, $TBP(X_3)$ (as a representative for $TBP(X_m)$ $(3 \leq m \leq n)$) might be computed via any of the formulas (5), (6) and (9)-(15). For example, formula (9) yields (thanks to (57) and (B.7))

$$TBP(X_3) = wt(\big(Sy(n-3;\{(k-1)\};\boldsymbol{X_r})\big)\,\overline{X}_1\overline{X}_2 \oplus \big(Sy(n-3;\{(k-2)\};\boldsymbol{X_r})\big)\,(\overline{X}_1X_2 \oplus X_1\overline{X}_2)) = c(n-3,k-1) + 2\,c(n-3,k-2). \quad (62)$$

In Appendix B, we show that the Public Good Index (PGI) $PGI(X_m)$ is equal to the binomial coefficient $c(n-1,k-1)$ (and hence to $TBP(X_m)$) for an unrestricted k-out-of-n system. If the two voters $X_1$ and $X_2$ refuse to share membership in any coalition, then to compute $PGI(X_m)$, we differentiate between the cases when $X_m$ is either $X_1$ or $X_2$, and the case it is neither of them. In the former case, we need to decrement the original value of $PGI(X_m)$ by the number of prime implicants in which the product $X_1X_2$ appears, which is the number of prime implicants of the $(k-2)$-out-of-$(n-2)$ function $Sy(n;\{k..n\};\boldsymbol{X})/X_1X_2 = Sy(n-2;\{(k-2)..(n-2)\};\boldsymbol{X_r})$, i.e., the binomial coefficient $c(n-2,k-2)$. This means that

$$PGI(X_1) = c(n-1,k-1) - c(n-2,k-2) = c(n-2,k-1) = TBP(X_1). \quad (63)$$

In the latter case, we need to decrement the original value of say $PGI(X_3)$ by the number of prime implicants in which the product $X_1X_2$ appears, and also $X_3$ appears, which is the number of prime implicants of the $(k-3)$-out-of-$(n-3)$ function $Sy(n;\{k..n\};\boldsymbol{X})/X_1X_2X_3 = Sy(n-3;\{(k-3)..(n-3)\};\boldsymbol{X_r}/X_3)$, i.e., the binomial coefficient $c(n-3,k-3)$. This means that

$$PGI(X_3) = c(n-1,k-1) - c(n-3,k-3) = c(n-3,k-1) + 2\,c(n-3,k-2) = TBP(X_3). \quad (64)$$

These results demonstrate that $PGI(X_m)$ is exactly equal to $TBP(X_m)$ for a general k-out-of-n system, not only when this system is unrestricted, but also when it is restricted through the lack of co-operation between two voting members. A visual demonstration of the above results can be obtained from the probability sample spaces of Figs. 3-5.

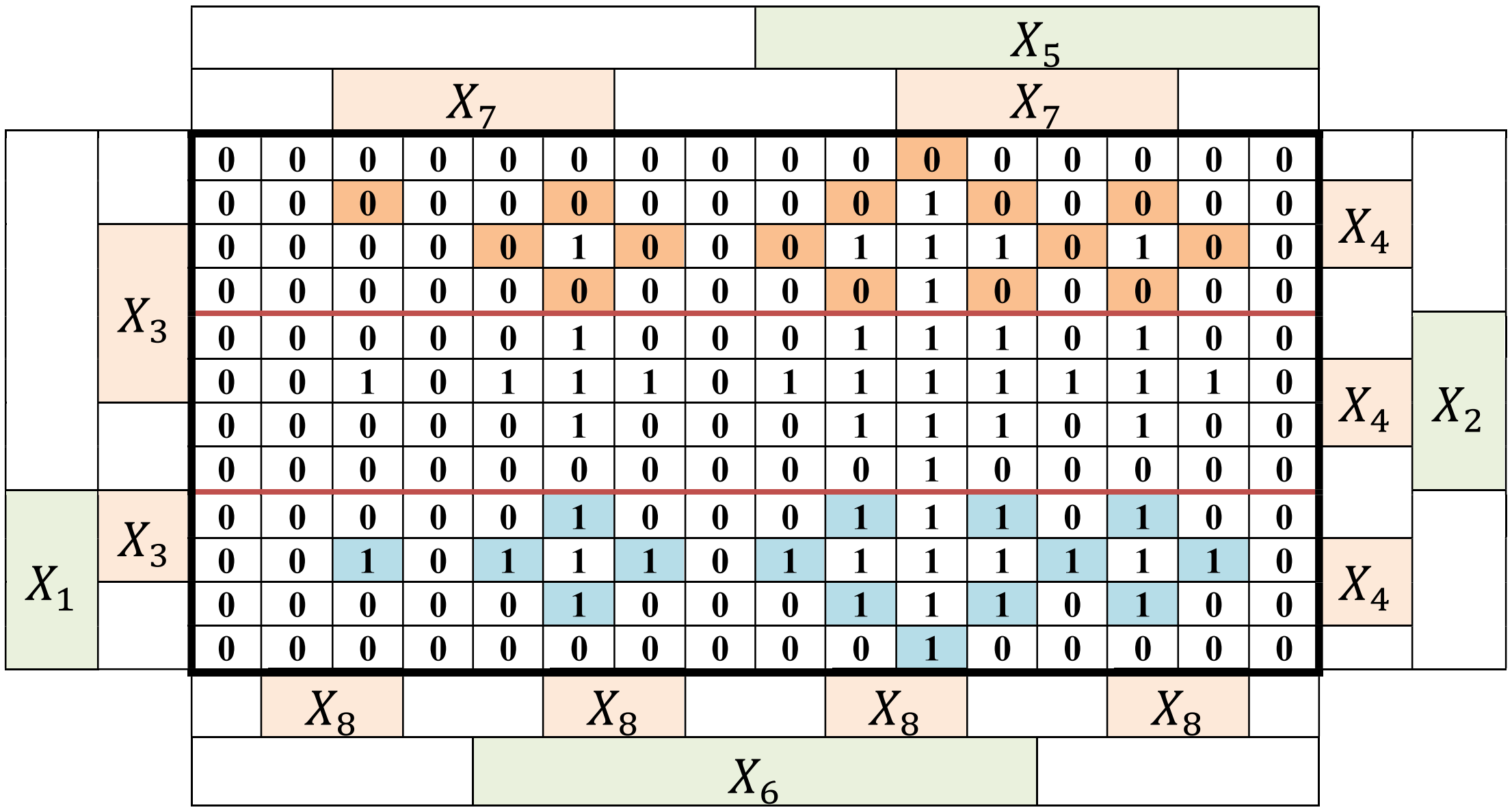

| | | | | | | | | | | | | | | | |
|---|---|---|---|---|---|---|---|---|---|---|---|---|---|---|---|
| 0 | 0 | 0 | 0 | 0 | 0 | 0 | 0 | 0 | 0 | 0 | 0 | 0 | 0 | 0 | 0 |
| 0 | 0 | 0 | 0 | 0 | 0 | 0 | 0 | 0 | 0 | 1 | 0 | 0 | 0 | 0 | 0 |
| 0 | 0 | 0 | 0 | 0 | 1 | 0 | 0 | 0 | 1 | 1 | 1 | 0 | 1 | 0 | 0 |
| 0 | 0 | 0 | 0 | 0 | 0 | 0 | 0 | 0 | 0 | 1 | 0 | 0 | 0 | 0 | 0 |
| 0 | 0 | 0 | 0 | 0 | 1 | 0 | 0 | 0 | 1 | 1 | 1 | 0 | 1 | 0 | 0 |
| 0 | 0 | 1 | 0 | 1 | 1 | 1 | 0 | 1 | 1 | 1 | 1 | 1 | 1 | 1 | 0 |
| 0 | 0 | 0 | 0 | 0 | 1 | 0 | 0 | 0 | 1 | 1 | 1 | 0 | 1 | 0 | 0 |
| 0 | 0 | 0 | 0 | 0 | 0 | 0 | 0 | 0 | 0 | 1 | 0 | 0 | 0 | 0 | 0 |
| 0 | 0 | 0 | 0 | 0 | 1 | 0 | 0 | 0 | 1 | 1 | 1 | 0 | 1 | 0 | 0 |
| 0 | 0 | 1 | 0 | 1 | 1 | 1 | 0 | 1 | 1 | 1 | 1 | 1 | 1 | 1 | 0 |
| 0 | 0 | 0 | 0 | 0 | 1 | 0 | 0 | 0 | 1 | 1 | 1 | 0 | 1 | 0 | 0 |
| 0 | 0 | 0 | 0 | 0 | 0 | 0 | 0 | 0 | 0 | 1 | 0 | 0 | 0 | 0 | 0 |

$g(\boldsymbol{X})$

Figure 3. The exact Universe of Discourse (probability sample space) that describes the new decision function $g(\boldsymbol{X})$ when the coalition $X_1X_2$ is not allowed in Example 2 for a 5-out-of-8 system. Here, the domain $X_1X_2 = 1$ is annihilated, $g(\boldsymbol{X})/X_1X_2$ does not exist, and statistical independence between $X_1$ and $X_2$ is lost. The sample space is a Karnaugh-map-like structure, but it is not a Karnaugh map *per se*. The function $g(\boldsymbol{X})$ ceases to be monotonically non-decreasing in each of $X_1$ and $X_2$ but continues to be so for each of the remaining variables. The map is partitioned into three 4-row sub-maps containing the Boolean quotients or subfunctions $g(\boldsymbol{X})/\overline{X}_1\overline{X}_2 = Sy(6; \{5..6\};\ X_3, X_4, X_5, X_6, X_7, X_8)$, and $g(\boldsymbol{X})/\overline{X}_1X_2 = g(\boldsymbol{X})/X_1\overline{X}_2 = Sy(6; \{4..6\};\ X_3, X_4, X_5, X_6, X_7, X_8)$. With the coalition $X_1X_2$ being forbidden, $TBP(X_1) = 15 = c(n-2, k-1) = c\,(6, 4)$ since there 15 solutions for the equation $(g(\boldsymbol{X})/X_1)\ (\overline{g}(\boldsymbol{X})/\overline{X}_1) = 1$, corresponding to the shown swings from a blue cell to an adjacent red one across the $X_1$ boundary. Note that for the unrestricted system, the original Banzhaf power index was $TBP(X_1) = 35 = c(n-1, k-1) = c\,(7, 4)$.

| | | | | | | | | | | | | | | | |
|---|---|---|---|---|---|---|---|---|---|---|---|---|---|---|---|
| 0 | 0 | 0 | 0 | 0 | 0 | 0 | 0 | 0 | 0 | 0 | 0 | 0 | 0 | 0 | 0 |
| 0 | 0 | 0 | 0 | 0 | 0 | 0 | 0 | 0 | 0 | 1 | 0 | 0 | 0 | 0 | 0 |
| 0 | 0 | 0 | 0 | 0 | 1 | 0 | 0 | 0 | 1 | 1 | 1 | 0 | 1 | 0 | 0 |
| 0 | 0 | 0 | 0 | 0 | 0 | 0 | 0 | 0 | 0 | 1 | 0 | 0 | 0 | 0 | 0 |
| 0 | 0 | 0 | 0 | 0 | 1 | 0 | 0 | 0 | 1 | 1 | 1 | 0 | 1 | 0 | 0 |
| 0 | 0 | 1 | 0 | 1 | 1 | 1 | 0 | 1 | 1 | 1 | 1 | 1 | 1 | 1 | 0 |
| 0 | 0 | 0 | 0 | 0 | 1 | 0 | 0 | 0 | 1 | 1 | 1 | 0 | 1 | 0 | 0 |
| 0 | 0 | 0 | 0 | 0 | 0 | 0 | 0 | 0 | 0 | 1 | 0 | 0 | 0 | 0 | 0 |
| 0 | 0 | 0 | 0 | 0 | 0 | 0 | 0 | 0 | 0 | 0 | 0 | 0 | 0 | 0 | 0 |
| 0 | 0 | 0 | 0 | 0 | 0 | 0 | 0 | 0 | 0 | 0 | 0 | 0 | 0 | 0 | 0 |
| 0 | 0 | 0 | 0 | 0 | 0 | 0 | 0 | 0 | 0 | 0 | 0 | 0 | 0 | 0 | 0 |
| 0 | 0 | 0 | 0 | 0 | 0 | 0 | 0 | 0 | 0 | 0 | 0 | 0 | 0 | 0 | 0 |
| 0 | 0 | 0 | 0 | 0 | 1 | 0 | 0 | 0 | 1 | 1 | 1 | 0 | 1 | 0 | 0 |
| 0 | 0 | 1 | 0 | 1 | 1 | 1 | 0 | 1 | 1 | 1 | 1 | 1 | 1 | 1 | 0 |
| 0 | 0 | 0 | 0 | 0 | 1 | 0 | 0 | 0 | 1 | 1 | 1 | 0 | 1 | 0 | 0 |
| 0 | 0 | 0 | 0 | 0 | 0 | 0 | 0 | 0 | 0 | 1 | 0 | 0 | 0 | 0 | 0 |

$g(\boldsymbol{X})$

Figure 4. The Karnaugh map for a restricted 5-out-of-8 system. When a coalition $X_1X_2$ is not allowed, it is mathematically convenient to obtain the new decision function $g(\boldsymbol{X})$ from the original one by forcing the yellow cells (of domain $X_1X_2 = 1$) to be 0's for $g(\boldsymbol{X})$. Again, the function $g(\boldsymbol{X})$ ceases to be monotonically non-decreasing in each of $X_1$ and $X_2$ but continues to be so for each of the remaining variables. The map is partitioned into four four-row sub-maps including the forbidden domain $X_1X_2 = 1$ (which is now restored (as compared to Fig. 3) and assigned 0's in its cells), $g(\boldsymbol{X})/\overline{X}_1\overline{X}_2 = Sy(6; \{5..6\};\ X_3, X_4, X_5, X_6, X_7, X_8)$, and $g(\boldsymbol{X})/\overline{X}_1X_2 = g(\boldsymbol{X})/X_1\overline{X}_2 = Sy(6; \{4..6\};\ X_3, X_4, X_5, X_6, X_7, X_8)$. With a coalition $X_1X_2$ being forbidden, $TBP(X_1) = 15$ since there 15 solutions for the equation $(g(\boldsymbol{X})/X_1)\ (\overline{g}(\boldsymbol{X})/\overline{X}_1) = 1$, corresponding to the shown swings from a blue cell to an adjacent red one across the $X_1$ boundary. Reverse swings are excluded from consideration through the utilization of the appropriate formulas (5) or (6).

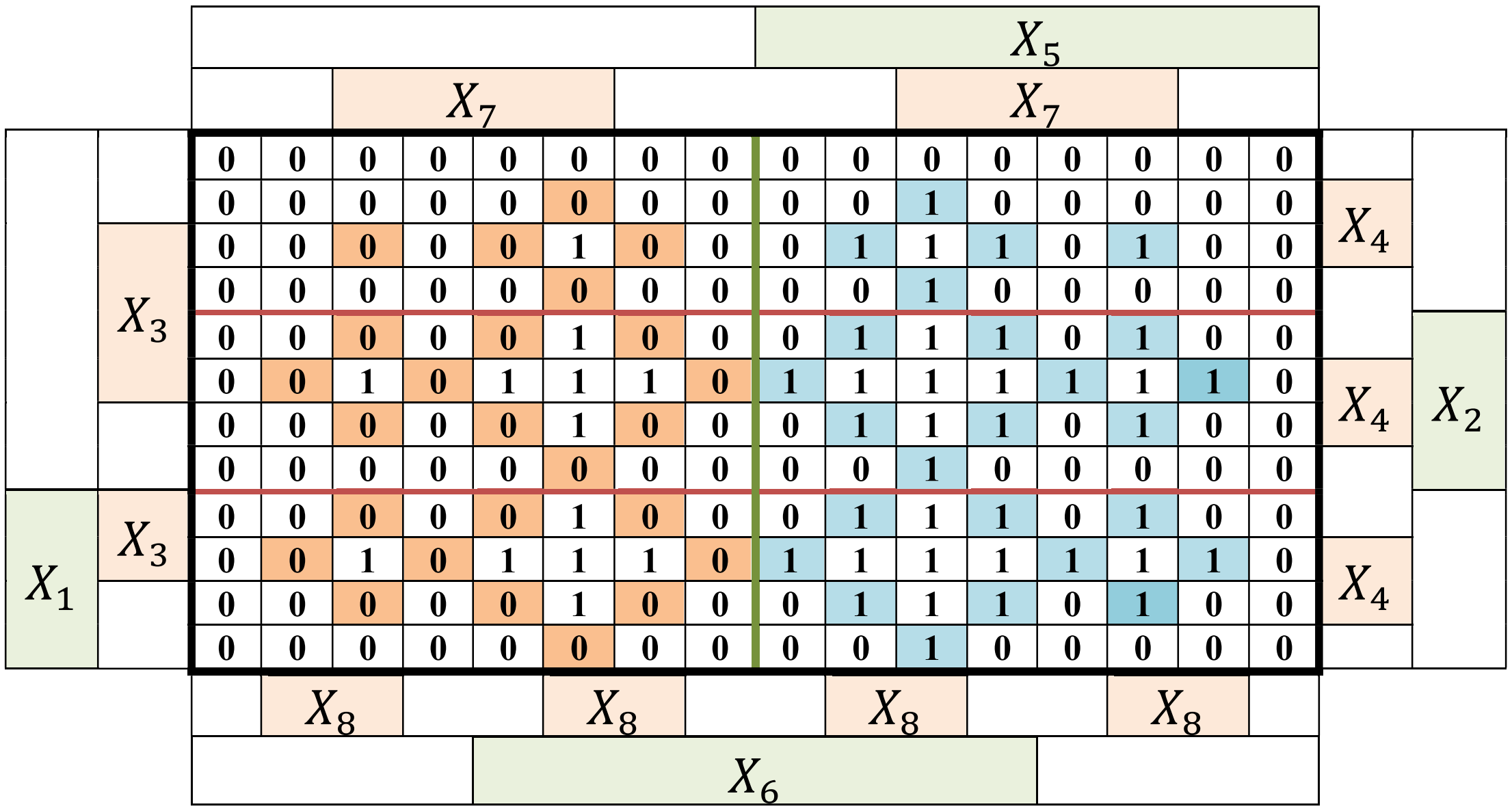

$g(\boldsymbol{X})$

Figure 5. The exact Universe of Discourse (probability sample space) for a restricted 5-out-of-8 system, which describes the new decision function $g(\boldsymbol{X})$ when a coalition $X_1X_2$ is not allowed in Example 2. Here, we show that $TBP(X_5) = 25 = c(n-3, k-1) + 2\,c(n-3, k-2) = c(5,4) + 2\,c(5,3)$, since there are 25 swings from a blue cell to an adjacent red one across the $X_5$ boundary. Denial of co-operation between $X_1$ and $X_2$ decreased $TBP(X_1) = TBP(X_2)$ more dramatically from 35 to 15, and also reduced $TBP(X_m)$ (for $3 \leq m \leq 8$) (albeit less dramatically) from 35 to 25.

## 5. A Five-Member Voting System

This section refers to the six-member voting system [65; 47, 46, 17, 16, 2, 1], which represents the Scottish Parliament of 2007 [12, 78]. The first five members of this system are political parties, while the sixth is an independent individual. This individual is somehow ignored in many studies (See, e.g., [77]) that view just the reduced five-member system [65; 47, 46, 17, 16, 2]. For simplicity, we will also consider the reduced system herein. For this 5-member system, the decision function $f(\boldsymbol{X})$ and its complement, which are both partially symmetric in $X_4$ and $X_5$ (despite the disparity between $W_4 = 16$ and $W_5 = 2$), are both readily available, respectively, as sums of disjoint products [12], namely

$$f(\boldsymbol{X}) = X_1X_2 \oplus X_1\overline{X}_2X_3X_4 \oplus X_1\overline{X}_2X_3\overline{X}_4X_5 \oplus X_1\overline{X}_2\overline{X}_3X_4X_5 \oplus \overline{X}_1X_2X_3X_4 \oplus \overline{X}_1X_2X_3\overline{X}_4X_5. \quad (65)$$

$$\overline{f}(\boldsymbol{X}) = \overline{X}_1\overline{X}_2 \oplus \overline{X}_1 X_2\overline{X}_3 \oplus \overline{X}_1 X_2 X_3 \overline{X}_4\overline{X}_5 \oplus X_1\overline{X}_2\overline{X}_3\overline{X}_4 \oplus X_1\overline{X}_2\overline{X}_3 X_4\overline{X}_5 \oplus X_1\overline{X}_2 X_3\overline{X}_4\overline{X}_5. \quad (66)$$

Now, we compute the various Banzhaf indices using the general formula (5)

$$TBP(X_1) = wt((f/X_1)(\overline{f}/\overline{X}_1)) = wt((X_2 \oplus \overline{X}_2 X_3 X_4 \oplus \overline{X}_2 X_3\overline{X}_4 X_5 \oplus \overline{X}_2\overline{X}_3 X_4 X_5)(\overline{X}_2 \oplus X_2\overline{X}_3 \oplus X_2 X_3 \overline{X}_4\overline{X}_5)) = wt(\overline{X}_2 X_3 X_4 \oplus \overline{X}_2 X_3\overline{X}_4 X_5 \oplus \overline{X}_2\overline{X}_3 X_4 X_5 \oplus X_2\overline{X}_3 \oplus X_2 X_3 \overline{X}_4\overline{X}_5) = 2 + 1 + 1 + 4 + 1 = 9. \quad (67)$$

$$TBP(X_2) = wt((f/X_2)(\overline{f}/\overline{X}_2)) = wt\left((X_1 \oplus \overline{X}_1 X_3 X_4 \oplus \overline{X}_1 X_3\overline{X}_4 X_5)(\overline{X}_1 \oplus X_1\overline{X}_3\overline{X}_4 \oplus X_1\overline{X}_3 X_4\overline{X}_5 \oplus X_1 X_3\overline{X}_4\overline{X}_5)\right) = wt(X_1\overline{X}_3\overline{X}_4 \oplus X_1\overline{X}_3 X_4\overline{X}_5 \oplus X_1 X_3\overline{X}_4\overline{X}_5 \oplus \overline{X}_1 X_3 X_4 \oplus \overline{X}_1 X_3\overline{X}_4 X_5) = 2 + 1 + 1 + 2 + 1 = 7. \quad (68)$$

$$TBP(X_3) = wt((f/X_3)(\overline{f}/\overline{X}_3)) = wt((X_1 X_2 \oplus X_1\overline{X}_2 X_4 \oplus X_1\overline{X}_2\overline{X}_4 X_5 \oplus \overline{X}_1 X_2 X_4 \oplus \overline{X}_1 X_2\overline{X}_4 X_5)(\overline{X}_1\overline{X}_2 \oplus \overline{X}_1 X_2 \oplus X_1\overline{X}_2\overline{X}_4 \oplus X_1\overline{X}_2 X_4\overline{X}_5) = wt(X_1\overline{X}_2 X_4\overline{X}_5 \oplus X_1\overline{X}_2\overline{X}_4 X_5 \oplus \overline{X}_1 X_2 X_4 \oplus \overline{X}_1 X_2\overline{X}_4 X_5 = 1 + 1 + 2 + 1 = 5. \quad (69)$$

$$TBP(X_4) = wt\left((f/X_4)(\overline{f}/\overline{X}_4)\right) = wt\left((X_1 X_2 \oplus X_1\overline{X}_2 X_3 \oplus X_1\overline{X}_2\overline{X}_3 X_5 \oplus \overline{X}_1 X_2 X_3)(\overline{X}_1\overline{X}_2 \oplus \overline{X}_1 X_2\overline{X}_3 \oplus \overline{X}_1 X_2 X_3 \overline{X}_5 \oplus X_1\overline{X}_2\overline{X}_3 \oplus X_1\overline{X}_2 X_3\overline{X}_5)\right) = wt(X_1\overline{X}_2 X_3\overline{X}_5 \oplus X_1\overline{X}_2\overline{X}_3 X_5 \oplus X_1\overline{X}_2 X_3\overline{X}_5) = 1 + 1 + 1 = 3. \quad (70)$$

$$TBP(X_5) = wt\left((f/X_5)(\overline{f}/\overline{X}_5)\right) = wt\left((X_1 X_2 \oplus X_1\overline{X}_2 X_3 X_4 \oplus X_1\overline{X}_2 X_3\overline{X}_4 \oplus X_1\overline{X}_2\overline{X}_3 X_4 \oplus \overline{X}_1 X_2 X_3 X_4 \oplus \overline{X}_1 X_2 X_3\overline{X}_4)(\overline{X}_1\overline{X}_2 \oplus \overline{X}_1 X_2\overline{X}_3 \oplus \overline{X}_1 X_2 X_3 \overline{X}_4 \oplus X_1\overline{X}_2\overline{X}_3\overline{X}_4 \oplus X_1\overline{X}_2\overline{X}_3 X_4 \oplus X_1\overline{X}_2 X_3\overline{X}_4)\right) = wt(X_1\overline{X}_2 X_3\overline{X}_4 \oplus X_1\overline{X}_2\overline{X}_3 X_4 \oplus \overline{X}_1 X_2 X_3\overline{X}_4) = 1 + 1 + 1 = 3. \quad (71)$$

Finally, the vector of total Banzhaf power is (the same as obtained in [12])

$$\boldsymbol{TBP} = [\,9 \quad 7 \quad 5 \quad 3 \quad 3]^T, \quad (72)$$

Now, we consider the case where the coalition $X_1 X_2$ is forbidden. We need to impose the constraint $\{X_1 X_2 = 0\}$ on the function $f(\boldsymbol{X})$, which leads to replacing $f(\boldsymbol{X})$ by another function, say $g(\boldsymbol{X})$. First, we construct the Boole-Shannon expansion of $f(\boldsymbol{X})$ w.r.t. the two variables $X_1$ and $X_2$, namely

$$f(\boldsymbol{X}) = \big(f(\boldsymbol{X})/\overline{X}_1\overline{X}_2\big)\,\overline{X}_1\overline{X}_2 \oplus \big(f(\boldsymbol{X})/\overline{X}_1X_2\big)\,\overline{X}_1X_2 \oplus \big(f(\boldsymbol{X})/X_1\overline{X}_2\big)\,X_1\overline{X}_2 \oplus \big(f(\boldsymbol{X})/X_1X_2\big)\,X_1X_2. \quad (73)$$

$$f(\boldsymbol{X}) = (0)\,\overline{X}_1\overline{X}_2 \oplus \big(X_3X_4 \oplus X_3\overline{X}_4X_5\big)\,\overline{X}_1X_2 \oplus \big(X_3X_4 \oplus X_3\overline{X}_4X_5 \oplus \overline{X}_3X_4X_5\big)\,X_1\overline{X}_2 \oplus (1)\,X_1X_2. \quad (74)$$

Now we construct $g(\boldsymbol{X})$ so as to share all subfunctions with $f(\boldsymbol{X})$, with the exception of the subfunction w.r.t. $X_1X_2$, which is nullified

$$g(\boldsymbol{X}) = (0)\,\overline{X}_1\overline{X}_2 \oplus \big(X_3X_4 \oplus X_3\overline{X}_4X_5\big)\,\overline{X}_1X_2 \oplus \big(X_3X_4 \oplus X_3\overline{X}_4X_5 \oplus \overline{X}_3X_4X_5\big)\,X_1\overline{X}_2 \oplus (\boldsymbol{0})\,X_1X_2 = X_1\overline{X}_2X_3X_4 \oplus X_1\overline{X}_2X_3\overline{X}_4X_5 \oplus X_1\overline{X}_2\overline{X}_3X_4X_5 \oplus \overline{X}_1X_2X_3X_4 \oplus \overline{X}_1X_2X_3\overline{X}_4X_5. \quad (75)$$

We obtain the complementary function $\overline{g}(\boldsymbol{X})$ by complementing each subfunction of $g(\boldsymbol{X})$, or by complementing each subfunction of $f(\boldsymbol{X})$ with the exception of the subfunction w.r.t. $X_1X_2$, which is now asserted

$$\overline{g}(\boldsymbol{X}) = (1)\,\overline{X}_1\overline{X}_2 \oplus \big(\overline{X}_3 \oplus X_3\,\overline{X}_4\,\overline{X}_5\big)\,\overline{X}_1X_2 \oplus \big(\overline{X}_3\overline{X}_4 \oplus \overline{X}_3\,X_4\overline{X}_5 \oplus X_3\overline{X}_4\overline{X}_5\big)\,X_1\overline{X}_2 \oplus (\boldsymbol{1})\,X_1X_2 = \overline{X}_1\overline{X}_2 \oplus \overline{X}_1\,X_2\overline{X}_3 \oplus \overline{X}_1X_2X_3\,\overline{X}_4\overline{X}_5 \oplus X_1\overline{X}_2\overline{X}_3\overline{X}_4 \oplus X_1\overline{X}_2\overline{X}_3\,X_4\overline{X}_5 \oplus X_1\overline{X}_2\,X_3\overline{X}_4\overline{X}_5 \oplus X_1X_2. \quad (76)$$

Note also that the functions $g(\boldsymbol{X})$ and $\overline{g}(\boldsymbol{X})$ are no longer unate; each of them is biform in $X_1$ and $X_2$, but each remains monoform in $X_3$, $X_4$ and $X_5$. The functions $g(\boldsymbol{X})$ and $\overline{g}(\boldsymbol{X})$ retain partial symmetry in $X_4$ and $X_5$, and hence $TBP(X_4) = TBP(X_5)$, and it suffices to compute one of these two values, say $TBP(X_4)$. Since $g(\boldsymbol{X})$ is no longer monotone in $X_1$ or $X_2$ , formulas (5) and (6) are the only valid formulas for computing $TBP(X_1)$ and $TBP(X_2)$, and we compute them via these two formulas as

$$TBP(X_1) = wt((g/X_1)(\overline{g}/\overline{X}_1)) = wt((\overline{X}_2X_3X_4 \oplus \overline{X}_2X_3\overline{X}_4X_5 \oplus \overline{X}_2\overline{X}_3X_4X_5)\,(\overline{X}_2 \oplus X_2\overline{X}_3 \oplus X_2X_3\,\overline{X}_4\overline{X}_5)) = wt(\overline{X}_2X_3X_4 \oplus \overline{X}_2X_3\overline{X}_4X_5 \oplus \overline{X}_2\overline{X}_3X_4X_5) = 2+1+1 = 4, \quad (77)$$

$$TBP(X_1) = 2^{n-1} - wt((\overline{g}/X_1) \vee (\,g/\overline{X}_1)) = 16 - wt(\big(\overline{X}_2\overline{X}_3\overline{X}_4 \oplus \overline{X}_2\overline{X}_3\,X_4\overline{X}_5 \oplus \overline{X}_2\,X_3\overline{X}_4\overline{X}_5 \oplus X_2\big) \vee (\,X_2X_3X_4 \oplus X_2X_3\overline{X}_4X_5)) = 16 - (2+1+1+8) - (2+1) + (2+1) = 4. \quad (78)$$

$$TBP(X_2) = wt((g/X_2)(\overline{g}/\overline{X}_2)) = wt((\,\overline{X}_1X_3X_4 \oplus \overline{X}_1X_3\overline{X}_4X_5)\,(\overline{X}_1 \oplus X_1\overline{X}_3\overline{X}_4 \oplus X_1\overline{X}_3\,X_4\overline{X}_5 \oplus X_1\,X_3\overline{X}_4\overline{X}_5)) = wt(\overline{X}_1X_3X_4 \oplus \overline{X}_1X_3\overline{X}_4X_5) = 2+1 = 3, \quad (79)$$

$$TBP(X_2) = 2^{n-1} - wt((\overline{g}/X_2) \vee (\,g/\overline{X}_2)) = 16 - wt(\big(\overline{X}_1\,\overline{X}_3 \oplus \overline{X}_1X_3\,\overline{X}_4\overline{X}_5 \oplus X_1\big) \vee (\,X_1X_3X_4 \oplus X_1X_3\overline{X}_4X_5 \oplus X_1\overline{X}_3X_4X_5)) = 16 - (4+1+8) - (2+1+1) + (2+1+1) = 3. \quad (80)$$

Now, we compute $TBP(X_3)$ via each of formulas (12) and (15) as

$$TBP(X_3) = wt(g/X_3) - wt(g/\overline{X}_3) = wt((X_1\overline{X}_2X_4 \oplus X_1\overline{X}_2\overline{X}_4X_5 \oplus \overline{X}_1X_2X_4 \oplus \overline{X}_1X_2\overline{X}_4X_5)) - wt(X_1\overline{X}_2X_4X_5) = (2+1+2+1) - 1 = 5. \quad (81)$$

$$TBP(X_3) = wt(\overline{g}/\overline{X}_3) - wt(\overline{g}/X_3)$$
$$= wt(\overline{X}_1\overline{X}_2 \oplus \overline{X}_1 X_2 \oplus X_1\overline{X}_2\overline{X}_4 \oplus X_1\overline{X}_2X_4\overline{X}_5 \oplus X_1X_2) -$$
$$wt(\overline{X}_1\overline{X}_2 \oplus \overline{X}_1X_2\overline{X}_4\overline{X}_5 \oplus X_1\overline{X}_2\overline{X}_4\overline{X}_5 \oplus X_1X_2) = (4+4+2+1+4) - (4+1+1+4) = 5, \quad (82)$$

We also compute $TBP(X_4) = TBP(X_5)$ via each of formulas (12) and (15) as

$$TBP(X_4) = wt(g/X_4) - wt(g/\overline{X}_4) = wt((X_1\overline{X}_2X_3 \oplus X_1\overline{X}_2\overline{X}_3X_5 \oplus \overline{X}_1X_2X_3)) - wt(X_1\overline{X}_2X_3X_5 \oplus \overline{X}_1X_2X_3X_5) = (2+1+2) - (1+1) = 3. \quad (83)$$

$$TBP(X_4) = wt(\overline{g}/\overline{X}_4) - wt(\overline{g}/X_4) = wt(\overline{X}_1\overline{X}_2 \oplus \overline{X}_1 X_2\overline{X}_3 \oplus \overline{X}_1X_2X_3\overline{X}_5 \oplus X_1\overline{X}_2\overline{X}_3 \oplus X_1\overline{X}_2X_3\overline{X}_5 \oplus X_1X_2) - wt(\overline{X}_1\overline{X}_2 \oplus \overline{X}_1X_2\overline{X}_3 \oplus X_1\overline{X}_2\overline{X}_3\overline{X}_5 \oplus X_1X_2) = (4+2+1+2+1+4) - (4+2+1+4) = 3. \quad (84)$$

Finally, the vector of total Banzhaf power when the coalition $X_1X_2$ is disallowed is

$$\boldsymbol{TBP} = [\,4 \quad 3 \quad 5 \quad 3 \quad 3]^T, \quad (85)$$

instead of its original value of $[\,9 \quad 7 \quad 5 \quad 3 \quad 3]^T$. It is interesting to note that in this example if the two largest parties refuse to cooperate, their powers decrease dramatically, while the remaining parties retain their original powers. In this restricted situation, the third largest party becomes the most powerful one.

For the present example, the Public Good Index (PGI) behaves similarly. Its value is $\boldsymbol{PGI} = [\,4 \quad 3 \quad 4 \quad 3 \quad 3]^T$ in the original case when all potential coalitions are allowed. If the coalition $X_1X_2$ is forbidden, this index simply changes to $\boldsymbol{PGI} = [\,3 \quad 2 \quad 4 \quad 3 \quad 3]^T$. The PGI again indicates a loss of power to those who refuse to form a coalition, and the rise of the third largest party to the status of the most powerful one. The fact that the Public Good Index (PGI) lacks local monotonicity with the weights (while the Banzhaf index does) is usually cited as a disadvantage for the PGI, but the present example perhaps suggests that this feature may be an asset for it in the very likely situation of a restriction on coalition formation.

## 6. A System Treated earlier without and with a Restriction

In this Section, we consider the [7; 4, 2, 1, 1, 1, 1, 1] voting system, considered earlier by Yakuba [33] without and with the restriction $X_2\, X_3 = 0$. Due to partial symmetries among the last five voters, it suffices to compute a voting measure for each of the first two voters $X_1$ and $X_2$, and for one only of the remaining five voters $X_3, X_4, X_5, X_6$, and $X_7$ (say $X_3$). With the restriction $X_2\, X_3 = 0$ being imposed, $X_3$ loses symmetry with the remaining four variables $X_4, X_5, X_6$, and $X_7$, but these four variables retain symmetry among themselves. In this restricted case, it suffices to compute a voting measure for each of the first three voters $X_1, X_2$ and $X_3$, and for one only of the remaining four voters $X_4, X_5, X_6$, and $X_7$ (say $X_4$).

First, we consider the decision function $f(\boldsymbol{X})$ of this system without any restriction and write its minimal sum (which happens to be also its complete sum since it is unate). For the function $f(\boldsymbol{X})$ to be asserted, there are three groups of possibilities, namely: (a) the support of the voters $X_1$ and $X_2$ together with at least one of the remaining five voters $X_3, X_4, X_5, X_6$, and $X_7$, (b) the support of the voter $X_1$ together with at least three of the last five voters, or (c) the support of the voter $X_2$ together with all the last five voters. Mathematically, this can be stated as

$$f(\boldsymbol{X}) = \; X_1 X_2 \; \mathrm{Sy}(5;\{1..5\}; X_3, X_4, X_5, X_6, X_7) \vee X_1 \; \mathrm{Sy}(5;\{3..5\}; X_3, X_4, X_5, X_6, X_7) \vee X_2 \; \mathrm{Sy}(5;\{5..5\}; X_3, X_4, X_5, X_6, X_7), \tag{86}$$

$$f(\boldsymbol{X}) = X_1 X_2 \; (X_3 \vee X_4 \vee X_5 \vee X_6 \vee X_7) \vee X_1 \, ( X_3 X_4 X_5 \vee X_3 X_4 X_6 \vee X_3 X_4 X_7 \vee X_3 X_5 X_6 \vee X_3 X_5 X_7 \vee X_3 X_6 X_7 \vee X_4 X_5 X_6 \vee X_4 X_5 X_7 \vee X_4 X_6 X_7 \vee X_5 X_6 X_7) \vee X_2 \; (X_3 X_4\, X_5 X_6 X_7). \tag{87}$$

The voter $X_1$ appears in all the 5 prime implicants of the first group and in all the 10 prime implicants of the second group. The voter $X_2$ appears in all the 5 prime implicants of the first group and in the single prime implicant of the third group. Each of the remaining five voters $X_3, X_4, X_5, X_6$, and $X_7$, appears in one of the 5 prime implicants of the first group and in 6 out of the 10 prime implicants of the second group, as well as in the single prime implicant of the third group. Therefore, the Public Good Index (PGI) is $\boldsymbol{PGI} = [15 \quad 6 \quad 8 \quad 8 \quad 8 \quad 8 \quad 8]^T$ in the original case when all potential coalitions are allowed. If the coalition $X_2 X_3$ is forbidden, this amounts to the deletion of the two prime implicants $X_1 X_2 X_3$ and $X_2 X_3 X_4\, X_5 X_6 X_7$, which amounts to a decrement by two for the PGI of $X_2$ and $X_3$ as well as a decrement by one for each of the other variables. This means that the Public Good Index simply changes to $\boldsymbol{PGI} = [14 \quad 4 \quad 6 \quad 7 \quad 7 \quad 7 \quad 7]^T$.

We now apply disjointing techniques [48, 51, 53, 54, 59, 60, 62] to obtain a disjoint sum for $f(\boldsymbol{X})$, in which the OR operator ($\vee$) can legitimately be replaced by the XOR operator ($\oplus$), namely

$$f(\boldsymbol{X}) = \ X_1X_2 \ \text{Sy}(5;\{1..5\};X_3,X_4,X_5,X_6,X_7) \ \oplus$$
$$X_1\overline{X}_2 \ \text{Sy}(5;\{3..5\};X_3,X_4,X_5,X_6,X_7) \oplus \overline{X}_1X_2 \ \text{Sy}(5;\{5..5\};X_3,X_4,X_5,X_6,X_7), \quad (88)$$

We now employ formulas (12) and (9) to obtain to the following total Banzhaf powers:

$$TBP(X_1) = wt(f/X_1) - wt(f/\overline{X}_1) = wt\Big(X_2 \ \text{Sy}(5;\{1..5\};X_3,X_4,X_5,X_6,X_7) \ \oplus$$
$$\overline{X}_2 \ \text{Sy}(5;\{3..5\};X_3,X_4,X_5,X_6,X_7)\Big) - wt\big(X_2 \ \text{Sy}(5;\{5..5\};X_3,X_4,X_5,X_6,X_7)\big) =$$
$$(1)(C(5,1) + C(5,3) - C(5,5)) = 31 + 16 - 1 = 46. \quad (89)$$

$$TBP(X_2) = wt(f/X_2) - wt(f/\overline{X}_2) = wt\Big( \ X_1 \ \text{Sy}(5;\{1..5\};X_3,X_4,X_5,X_6,X_7) \ \oplus$$
$$\overline{X}_1 \ \text{Sy}(5;\{5..5\};X_3,X_4,X_5,X_6,X_7)\Big) - wt\big( \ X_1 \ \text{Sy}(5;\{3..5\};X_3,X_4,X_5,X_6,X_7)\big) =$$
$$C(5{,}1) + C(5{,}5) - C(5{,}3) = 31 + 1 - 16 = 16. \quad (90)$$

$$TBP(X_3) = wt\left(\frac{\partial f}{\partial X_3}\right) = (1)\big(c(4,0) + c(4,2) + c(4,4)\big) = 1 + 6 + 1 = 8. \quad (91)$$

Therefore, the vector of total Banzhaf power under no restriction is (the same as obtained in [33])

$$\boldsymbol{TBP} = [46 \quad 16 \quad 8 \quad 8 \quad 8 \quad 8 \quad 8]^T, \quad (92)$$

Now, we consider the case where the coalition $X_2X_3$ is forbidden. We need to impose the constraint $\{X_2X_3 = 0\}$ on the function $f(\boldsymbol{X})$, which leads to replacing $f(\boldsymbol{X})$ by another function, say $g(\boldsymbol{X})$. First, we construct the Boole-Shannon expansion of $f(\boldsymbol{X})$ w.r.t. the two variables $X_2$ and $X_3$, namely

$$f(\boldsymbol{X}) = \big(f(\boldsymbol{X})/\overline{X}_2\overline{X}_3\big)\,\overline{X}_2\overline{X}_3 \oplus \big(f(\boldsymbol{X})/\overline{X}_2X_3\big)\,\overline{X}_2X_3 \oplus \big(f(\boldsymbol{X})/X_2\overline{X}_3\big)\,X_2\overline{X}_3 \oplus$$
$$(f(\boldsymbol{X})/X_2X_3)\,X_2X_3. \quad (93)$$

$$f(\boldsymbol{X}) = \big(X_1 \ \text{Sy}(4;\{3..4\};X_4,X_5,X_6,X_7)\big)\,\overline{X}_2\overline{X}_3 \ \oplus$$
$$\big( \ X_1 \ \text{Sy}(4;\{2..4\};X_4,X_5,X_6,X_7)\big)\,\overline{X}_2X_3 \oplus \big(X_1 \ \text{Sy}(4;\{1..4\};X_4,X_5,X_6,X_7)\big)\,X_2\overline{X}_3 \ \oplus$$
$$(X_1 \ \oplus \ \overline{X}_1 \ \text{Sy}(4;\{4..4\};X_4,X_5,X_6,X_7))\,X_2X_3. \quad (94)$$

Now we construct $g(\boldsymbol{X})$ so as to share all subfunctions with $f(\boldsymbol{X})$, with the exception of the subfunction w.r.t. $X_2X_3$, which is nullified

$$g(\boldsymbol{X}) = \big(X_1 \ \text{Sy}(4;\{3..4\};X_4,X_5,X_6,X_7)\big)\,\overline{X}_2\overline{X}_3 \oplus$$
$$\big( \ X_1 \ \text{Sy}(4;\{2..4\};X_4,X_5,X_6,X_7)\big)\,\overline{X}_2X_3 \oplus \big( \ X_1 \ \text{Sy}(4;\{1..4\};X_4,X_5,X_6,X_7)\big)\,X_2\overline{X}_3 \oplus$$
$$(\boldsymbol{0})\,X_2X_3. \quad (95)$$

We obtain the complementary function $\overline{g}(\boldsymbol{X})$ by complementing each subfunction of $g(\boldsymbol{X})$, or by complementing each subfunction of $f(\boldsymbol{X})$ apart from the subfunction w.r.t. $X_2X_3$, which is now asserted

$$\overline{g}(\boldsymbol{X}) = \left(\overline{X}_1 \oplus X_1 \,\mathrm{Sy}(4;\{0..2\};X_4,X_5,X_6,X_7)\right) \overline{X}_2\overline{X}_3 \oplus \left(\overline{X}_1 \oplus X_1 \,\mathrm{Sy}(4\{0..1\};X_4,X_5,X_6,X_7)\right) \overline{X}_2X_3 \oplus \left(\overline{X}_1 \oplus X_1 \,\mathrm{Sy}(4;\{0\};X_4,X_5,X_6,X_7)\right) X_2\overline{X}_3 \oplus (\mathbf{1})\, X_2X_3. \quad (96)$$

Note also that the functions $g(\boldsymbol{X})$ and $\overline{g}(\boldsymbol{X})$ are no longer unate; each of them is biform in $X_2$ and $X_3$, but it remains monoform in the remaining variables. Since $g(\boldsymbol{X})$ is no longer monotone in $X_2$ and $X_3$, formulas (5) and (6) are the only valid formulas for computing $TBP(X_2)$ and $TBP(X_3)$. We use formula (5) to compute these two values as follows:

$$\begin{aligned} TBP(X_2) &= wt((g/X_2)(\overline{g}/\overline{X}_2)) \\ &= wt\,(\left(X_1 \,\mathrm{Sy}(4;\{1..4\};X_4,X_5,X_6,X_7)\right)\overline{X}_3)(\left(\overline{X}_1 \right. \\ &\quad \left. \oplus X_1 \,\mathrm{Sy}(4;\{0..2\};X_4,X_5,X_6,X_7)\right) \overline{X}_3 \\ &\quad \oplus \left(\overline{X}_1 \oplus X_1 \,\mathrm{Sy}(4;\{0..1\};X_4,X_5,X_6,X_7)\right) X_3)) \end{aligned}$$

$$TBP(X_2) = wt\left(\left(X_1 \,\mathrm{Sy}(4;\{1..2\};X_4,X_5,X_6,X_7)\right)\overline{X}_3\right) = c(4,1) + c(4,2) = 4 + 6 = 10. \quad (97)$$

$$TBP(X_3) = wt((g/X_3)(\overline{g}/\overline{X}_3)) = wt(((X_1 \,\mathrm{Sy}(4;\{2..4\};X_4,X_5,X_6,X_7))\overline{X}_2)\,((\overline{X}_1 \oplus X_1 \,\mathrm{Sy}(4;\{0..2\};X_4,X_5,X_6,X_7))\,\overline{X}_2 \oplus \left(\overline{X}_1 \oplus X_1 \,\mathrm{Sy}(4;\{0\};X_4,X_5,X_6,X_7)\right) X_2)) = wt((X_1 \,\mathrm{Sy}(4;\{2\};X_4,X_5,X_6,X_7))\overline{X}_2) = c(4,2) = 6. \quad (98)$$

We also use formulas (12) and (9) to compute $TBP(X_1)$ and $TBP(X_4)$ as follows:

$$TBP(X_1) = wt(g/X_1) - wt(g/\overline{X}_1) = wt\left(\left(\mathrm{Sy}(4;\{3..4\};X_4,X_5,X_6,X_7)\right)\overline{X}_2\overline{X}_3 \oplus \left(\mathrm{Sy}(4;\{2..4\};X_4,X_5,X_6,X_7)\right)\overline{X}_2X_3 \oplus \left(\mathrm{Sy}(4;\{1..4\};X_4,X_5,X_6,X_7)\right)X_2\overline{X}_3\right) - wt(0) = (1)\left(C(4,3) + C(4,2) + C(4,1)\right) - 0 = 5 + 11 + 15 = 31. \quad (99)$$

$$TBP(X_4) = wt\left(\frac{\partial g}{\partial X_4}\right) = (1)\left(c(3,\ 2) + c(3,1) + c(3,0)\right) = 3 + 3 + 1 = 7. \quad (100)$$

Throughout this paper, we systematically employed the Boole-Shannon expansion to derive the restricted function $g(\boldsymbol{X})$. Occasionally, we can resort to some *ad hoc* manipulation to obtain a shortcut, attain some simplicity, and save some effort. For our present example, we can rewrite $f(\boldsymbol{X})$ as

$$f(\boldsymbol{X}) = X_1X_2\,\left(X_3 \oplus \overline{X}_3\right)\mathrm{Sy}(5;\{1..5\};X_3,X_4,X_5,X_6,X_7) \oplus X_1\overline{X}_2\,\mathrm{Sy}(5;\{3..5\};X_3,X_4,X_5,X_6,X_7) \oplus \overline{X}_1X_2\,(X_3X_4\,X_5X_6X_7), \quad (101)$$

and then obtain $g(\boldsymbol{X})$ by imposing $X_2X_3 = 0$ in the first and third groups of terms and replacing $\overline{X}_3\,\mathrm{Sy}(5;\{1..5\};X_3,X_4,X_5,X_6,X_7)$ by $\overline{X}_3\,\mathrm{Sy}(4;\{1..4\};X_4,X_5,X_6,X_7)$ to obtain the simpler results

$$g(\boldsymbol{X}) = X_1 X_2 \overline{X}_3 \operatorname{Sy}(4; \{1..4\}; X_4, X_5, X_6, X_7) \oplus X_1 \overline{X}_2 \operatorname{Sy}(5; \{3..5\}; X_3, X_4, X_5, X_6, X_7), \quad (102)$$

$$\overline{g}(\boldsymbol{X}) = X_1 X_2 \overline{X}_3 \operatorname{Sy}(4; \{0\}; X_4, X_5, X_6, X_7) \oplus X_1 \overline{X}_2 \operatorname{Sy}(5; \{0..2\}; X_3, X_4, X_5, X_6, X_7). \quad (103)$$

$$TBP(X_2) = wt((g/X_2)(\overline{g}/\overline{X}_2)) = wt((X_1 \overline{X}_3 \operatorname{Sy}(4; \{1..4\}; X_4, X_5, X_6, X_7))( X_1 \operatorname{Sy}(5; \{0..2\}; X_3, X_4, X_5, X_6, X_7))) = wt(X_1 \overline{X}_3 \operatorname{Sy}(4; \{1, 2\}; X_4, X_5, X_6, X_7)) = (1)\,(c(4,1) + c(4,2)) = 4 + 6 = 10. \quad (104)$$

$$TBP(X_3) = wt((g/X_3)(\overline{g}/\overline{X}_3)) = wt(( X_1 \overline{X}_2 \operatorname{Sy}(4; \{2..4\}; X_4, X_5, X_6, X_7))(X_1 X_2 \operatorname{Sy}(4; \{0\}; X_4, X_5, X_6, X_7) \oplus X_1 \overline{X}_2 \operatorname{Sy}(4; \{0..2\}; X_4, X_5, X_6, X_7))) = wt( X_1 \overline{X}_2 \operatorname{Sy}(4; \{2\}; X_4, X_5, X_6, X_7) = (1)\, c(4,2) = 6. \quad (105)$$

$$TBP(X_1) = wt(g/X_1) - wt\big(g/\overline{X}_1\big) = wt(X_2 \overline{X}_3 \operatorname{Sy}(4; \{1..4\}; X_4, X_5, X_6, X_7) \oplus \overline{X}_2 \operatorname{Sy}(5; \{3..5\}; X_3, X_4, X_5, X_6, X_7)) - wt(0) = C(4,1) + C(5,3) = 15 + 16 = 31. \quad (106)$$

$$TBP(X_4) = wt\left(\frac{\partial g}{\partial X_4}\right) = (1)\, c(3,0) + (1)\, c(4,2) = 1 + 6 = 7. \quad (107)$$

Finally, the vector of total Banzhaf power under the restriction ($X_2 X_3 = 0$) is (the same as obtained in [33])

$$\boldsymbol{TBP} = [31 \quad 10 \quad 6 \quad 7 \quad 7 \quad 7 \quad 7]^T. \quad (108)$$

The power of voter 1 (the most powerful voter) reduced to 31/46 = 67.4% of its original value. Meanwhile, the powers of voters 2 and 3 (the ones who refused to share a coalition) reduced respectively to 10/16 = 62.5% and 6/8 = 75% of their original values. The power for each of the remaining voters experienced the least relative reduction of 7/8 = 87.5%. This means that the power of every voter decreased, though the decrease is more pronounced for voters refusing to share a coalition and also for more powerful voters.

A visual demonstration of the above results can be obtained from the probability sample spaces of Figs. 6-8.

| | | | | | | | | | | | |
|---|---|---|---|---|---|---|---|---|---|---|---|
| | | | | | | $X_5$ | | | | | |
| | | | $X_7$ | | | | $X_7$ | | | | |
| | | 0 | 0 | 0 | 0 | 0 | 0 | 0 | 0 | | |
| | | 0 | 0 | 0 | 0 | 0 | 0 | 0 | 0 | $X_4$ | |
| | $X_3$ | 0 | 0 | 0 | 0 | 0 | 0 | 0 | 0 | | |
| | | 0 | 0 | 0 | 0 | 0 | 0 | 0 | 0 | | |
| | | 0 | 0 | 0 | 0 | 0 | 0 | 0 | 0 | | $X_2$ |
| | | 0 | 0 | 0 | 0 | 0 | 1 | 0 | 0 | $X_4$ | |
| | | 0 | 0 | 0 | 0 | 0 | 0 | 0 | 0 | | |
| | | 0 | 0 | 0 | 0 | 0 | 0 | 0 | 0 | | |
| $X_1$ | | 0 | 1 | 1 | 1 | 1 | 1 | 1 | 1 | | |
| | | 1 | 1 | 1 | 1 | 1 | 1 | 1 | 1 | $X_4$ | |
| | $X_3$ | 1 | 1 | 1 | 1 | 1 | 1 | 1 | 1 | | |
| | | 1 | 1 | 1 | 1 | 1 | 1 | 1 | 1 | | |
| | | 0 | 0 | 1 | 0 | 1 | 1 | 1 | 0 | | |
| | | 0 | 1 | 1 | 1 | 1 | 1 | 1 | 1 | $X_4$ | |
| | | 0 | 0 | 1 | 0 | 1 | 1 | 1 | 0 | | |
| | | 0 | 0 | 0 | 0 | 0 | 1 | 0 | 0 | | |
| | | | | $X_6$ | | | | | | | |

$$f(\boldsymbol{X})$$

Figure 6. Karnaugh map for the decision function of the unrestricted [7; 4, 2, 1, 1, 1, 1, 1] voting system. The Banzhaf index $TBP(X_1) = 46$ since there are 46 swings from a blue cell to an adjacent red one across the $X_1$ boundary.

| | | | | | | | | | | | |
|---|---|---|---|---|---|---|---|---|---|---|---|
| | | | | | | | $X_5$ | | | | |
| | | | $X_7$ | | | | $X_7$ | | | | |
| | | 0 | 0 | 0 | 0 | 0 | 0 | 0 | 0 | | |
| | | 0 | 0 | 0 | 0 | 0 | 0 | 0 | 0 | $X_4$ | |
| | $X_3$ | 0 | 0 | 0 | 0 | 0 | 0 | 0 | 0 | | |
| | | 0 | 0 | 0 | 0 | 0 | 0 | 0 | 0 | | |
| | | 0 | 0 | 0 | 0 | 0 | 0 | 0 | 0 | | |
| | | 0 | 0 | 0 | 0 | 0 | 1 | 0 | 0 | $X_4$ | |
| | | 0 | 0 | 0 | 0 | 0 | 0 | 0 | 0 | | |
| | | 0 | 0 | 0 | 0 | 0 | 0 | 0 | 0 | | $X_2$ |
| | | 0 | 1 | 1 | 1 | 1 | 1 | 1 | 1 | | |
| | | 1 | 1 | 1 | 1 | 1 | 1 | 1 | 1 | $X_4$ | |
| | | 0 | 0 | 0 | 0 | 0 | 0 | 0 | 0 | | |
| $X_1$ | $X_3$ | 0 | 0 | 0 | 0 | 0 | 0 | 0 | 0 | | |
| | | 0 | 0 | 1 | 0 | 1 | 1 | 1 | 0 | | |
| | | 0 | 1 | 1 | 1 | 1 | 1 | 1 | 1 | $X_4$ | |
| | | 0 | 0 | 1 | 0 | 1 | 1 | 1 | 0 | | |
| | | 0 | 0 | 0 | 0 | 0 | 1 | 0 | 0 | | |
| | | | | | $X_6$ | | | | | | |

$$g(\boldsymbol{X})$$

Figure 7. Karnaugh map for the decision function of the [7; 4, 2, 1, 1, 1, 1, 1] voting system, with $X_2X_3 = 0$. Instead of omitting the domain $X_2X_3 = 1$, we forced its cells (colored yellow) to contain 0's. The Banzhaf index of the first voter (the one with the largest weight) $TBP(X_1)$ reduced from 46 in Fig. 6 to 31 herein since there are now 31 swings from a blue cell to an adjacent red one across the $X_1$ boundary. This power is significantly impaired (even more than that of $X_3$, which is one of that voters that participated in causing the restriction).

$X_5$ (columns 5–8); $X_7$ (columns 2–3 and 6–7); $X_6$ (columns 3–6); $X_1$ (rows 9–16); $X_2$ (rows 5–12); $X_3$ (rows 3–6 and 11–14); $X_4$ (rows 2–3, 6–7, 10–11, 14–15)

| | | | | | | | |
|---|---|---|---|---|---|---|---|
| 0 | 0 | 0 | 0 | 0 | 0 | 0 | 0 |
| 0 | 0 | 0 | 0 | 0 | 0 | 0 | 0 |
| 0 | 0 | 0 | 0 | 0 | 0 | 0 | 0 |
| 0 | 0 | 0 | 0 | 0 | 0 | 0 | 0 |
| 0 | 0 | 0 | 0 | 0 | 0 | 0 | 0 |
| 0 | 0 | 0 | 0 | 0 | 1 | 0 | 0 |
| 0 | 0 | 0 | 0 | 0 | 0 | 0 | 0 |
| 0 | 0 | 0 | 0 | 0 | 0 | 0 | 0 |
| 0 | 1 | 1 | 1 | 1 | 1 | 1 | 1 |
| 1 | 1 | 1 | 1 | 1 | 1 | 1 | 1 |
| 0 | 0 | 0 | 0 | 0 | 0 | 0 | 0 |
| 0 | 0 | 0 | 0 | 0 | 0 | 0 | 0 |
| 0 | 0 | 1 | 0 | 1 | 1 | 1 | 0 |
| 0 | 1 | 1 | 1 | 1 | 1 | 1 | 1 |
| 0 | 0 | 1 | 0 | 1 | 1 | 1 | 0 |
| 0 | 0 | 0 | 0 | 0 | 1 | 0 | 0 |

$g(\boldsymbol{X})$

Figure 8. The Karnaugh map of Fig. 7 redrawn to demonstrate that the Banzhaf index $TBP(X_2) = 10$ since there are 10 swings from a blue cell to an adjacent red one across the $X_2$ boundary.

## 7. Discussions

We note that restrictions on coalition formation generally arise due to the presence of several sets of several voters each with these voters refusing to cooperate, i.e., with these voters rejecting any proposal to be together in the same coalition. For the sake of mathematical simplicity, and as a first cut at the problem, we limited our present exploration to the particular case when there is exactly a single set of exactly two voters who refuse to cooperate. This case might be partitioned into two sub-cases:

(a) The case when the two voters considered constitute a minimal winning coalition (MWC), or a prime implicant of the unrestricted decision function $f(\boldsymbol{X})$. Here, our examples in Sections 3 and 5 show (according to both the Banzhaf Index and the Public Good Index) that coalition restriction results in a decrease in the voting powers of the two non-cooperating voters, but produces no effect on the voting powers of the other voters.

(b) The case when the two voters considered do not constitute a specific MWC, but they are parts of several MWCs. Here, our examples in

Sections 4 and 6 show (again according to both the Banzhaf Index and the Public Good Index) that coalition restriction results in a decrease in the voting powers of all voters, but this decrease is more pronounced for highly-weighted voters and for the two non-cooperating voters.

Beside the qualitative similarity between the predictions of the Banzhaf Index and the Public Good Index for systems subject to restrictions on coalition formation, there is a quantitative agreement between these two indexes for a broad class of monotone voting systems, namely, the k-out-of-n voting systems. For these systems (whether they are unrestricted or restricted through denial of cooperation between two voters), the two indexes exactly agree in numerical value.

We now consider the generalization of the subject of exploration from one of a restriction on the formation of a specific coalition between exactly two voters to one of several restrictions on the formation of coalitions that involve several voters each. In this latter case, we construct the Boole-Shannon expansion w.r.t. all the variables that appear in any of the forbidden coalitions. If there are $k$ such variables, then the Boole-Shannon expansion comprises $2^k$ Boolean quotients. We then convert the decision function $f(\boldsymbol{X})$ into a restricted one $g(\boldsymbol{X})$ by nullifying each of the Boolean quotients w.r.t. any product that subsumes at least one of the forbidden coalitions, while leaving the remaining Boolean quotients intact.

As an example, Fig. 9 displays the Karnaugh map for an unspecified eight-member voting system with voters $(X_1, X_2, X_3, X_4, X_5, X_6, X_7, X_8)$ under the restrictions that neither of the three coalitions $X_1X_2$, $X_1X_5$, and $X_5X_6$ can be formed. We construct the Boole-Shannon expansion w.r.t. all the variables that appear in at least one of the three forbidden coalitions, namely the four variables $X_1, X_2, X_5$, and $X_6$. This expansion corresponds to partitioning the map into the 16 sub-maps shown. Next, we nullify the eight yellow-pasted domains $X_1X_2X_5X_6$, $X_1X_2X_5\overline{X}_6$, $X_1X_2\overline{X}_5X_6$, $X_1X_2\overline{X}_5\overline{X}_6$, $X_1\overline{X}_2X_5X_6$, $X_1\overline{X}_2X_5\overline{X}_6$, $\overline{X}_1X_2X_5X_6$, and $\overline{X}_1\overline{X}_2X_5X_6$, where each of these eight products subsumes at least one of the forbidden products $X_1X_2$, $X_1X_5$, and $X_5X_6$. Actually, the eight nullified domains do not really exist in the exact sample space. Now, we compute the total Banzhaf power of any of the four variables $X_1, X_2, X_5$, and $X_6$ by employing either formula (5) or formula (6). As for the four remaining variables $X_3, X_4, X_7$, and $X_8$, we have the liberty to compute the total Banzhaf power by any of the formulas (5), (6), or (9)-(15).

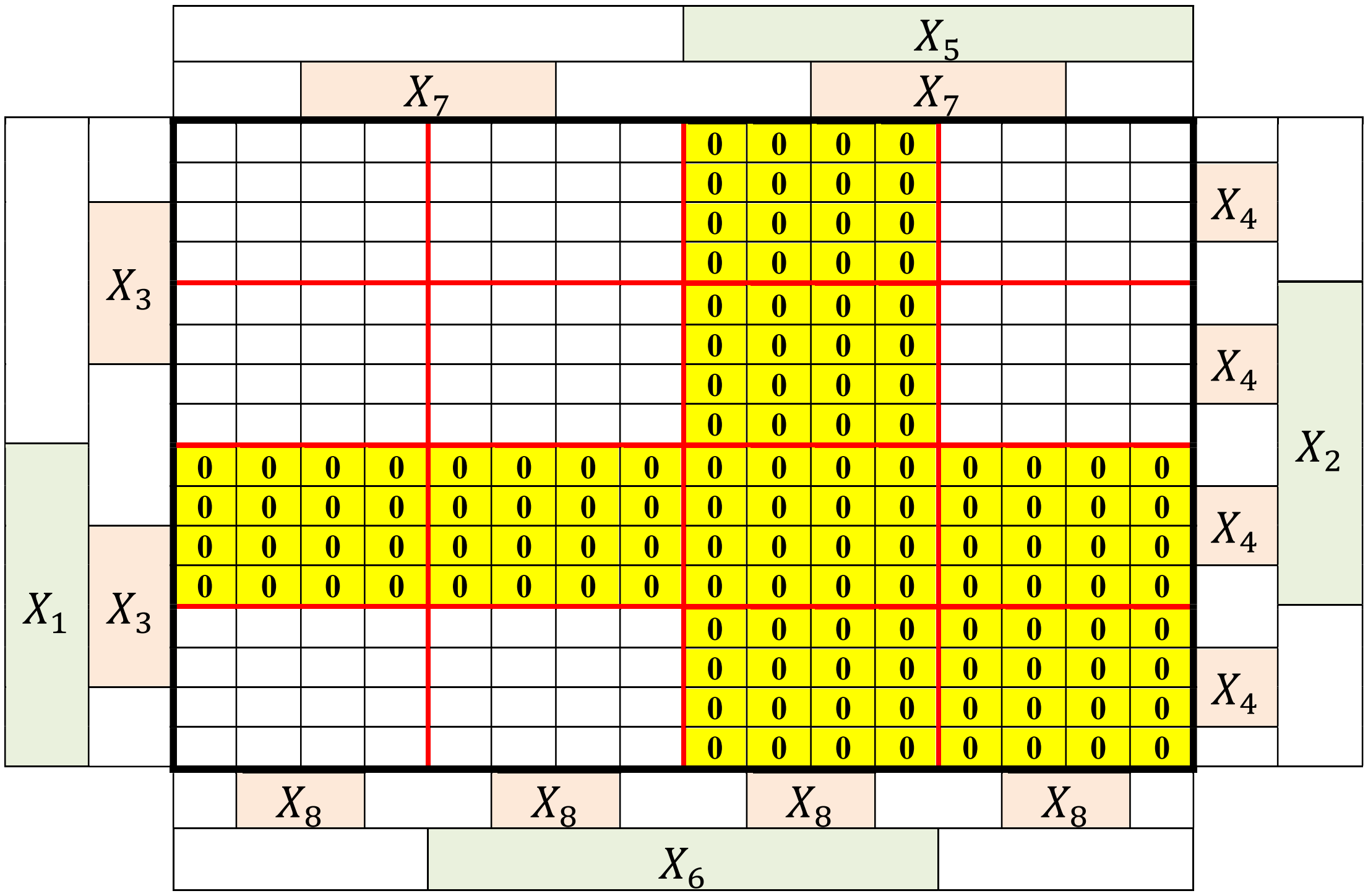

$f(\boldsymbol{X})$

Figure 9. The Karnaugh map for an unspecified eight-member voting system with voters $(X_1, X_2, X_3, X_4, X_5, X_6, X_7, X_8)$ under the restrictions that coalitions $X_1X_2$, $X_1X_5$, and $X_5X_6$ cannot be formed. We construct the Boole-Shannon expansion w.r.t. the four variables $X_1, X_2, X_5$, and $X_6$, (corresponding to partitioning the map into the 16 sub-maps shown) and then nullify the eight pasted-yellow domains $X_1X_2X_5X_6$, $X_1X_2X_5\overline{X}_6$, $X_1X_2\overline{X}_5X_6$, $X_1X_2\overline{X}_5\overline{X}_6$, $X_1\overline{X}_2X_5X_6$, $X_1\overline{X}_2X_5\overline{X}_6$, $\overline{X}_1X_2X_5X_6$, and $\overline{X}_1\overline{X}_2X_5X_6$. Each of these eight products subsumes at least one of the products $X_1X_2$, $X_1X_5$, and $X_5X_6$. Actually, the eight nullified domains do not really exist in the exact sample space.

## 8. Conclusions

This paper is another major step within our ongoing project [12, 19, 21, 22, 28-30] that strives to construct a switching-algebraic theory for weighted voting systems. This forthcoming theory is naturally expected to be a complement, a supplement, or an enhancement rather than a replacement or a substitute of the more dominant and all-encompassing game-theoretic description of these systems. The paper achieved its purpose through the exploration of a prominent index of voting power, *viz.,* the Banzhaf Index, as well as some related indices, such as the Public Good Index (PGI), with a stress on the case of a restriction on coalition formation. The paper attempts to make the most of the notable switching-algebraic concept of Boolean quotients. Detailed, multiple, and visual solutions are offered for four prominent examples, each of which covered both unrestricted and restricted voting systems. Finally, we reiterate from [12] that we believe that after completing a few planned-for tasks, we will be able to tell the entire tale of voting systems from a purely switching-algebraic perspective.

## Appendix A: Boolean Quotients

In this appendix, we deal with switching (two-valued Boolean) functions of the form $f(\boldsymbol{X}) = f(X_1, X_2, \dots, X_{m-1}, X_m, X_{m+1}, \dots, X_n)$. We define a literal to be a letter or its complement, where a letter is a constant or a variable (which could be one of the aforementioned arguments $X_m$, or an independent variable $Y$). A Boolean term or product is a conjunction or ANDing of $m$ literals in which no letter appears more than once (no letter is irredundant). For $m = 1$, a term is a single literal and for $m = 0$, a term is the constant 1. Note that, according to this definition the constant 0 is not a term. Given a Boolean function $f(\boldsymbol{X})$ and a term t, the Boolean quotient of $f$ with respect to t, denoted by $(f(\boldsymbol{X})/t)$, is defined to be the function formed from $f(\boldsymbol{X})$ by imposing the constraint $\{t = 1\}$ explicitly on it [12, 30, 44-52], i.e.,

$$f(\boldsymbol{X})/t = [f(\boldsymbol{X})]_{t=1}, \qquad (A.1)$$

For example, if $t = X_2\bar{X}_m$ then $f(\boldsymbol{X})/t = f(X_1, 1, \dots, X_{m-1}, 0, X_{m+1}, \dots, X_n)$, simply obtained by imposing $X_2\bar{X}_m = 1$, or substituting $X_2 = 1$ and $X_m = 0$ in $f(\boldsymbol{X})$. The Boolean quotient is also known as a ratio [79, 80], a subfunction [48, 52, 58, 81-83], or a restriction [12, 50, 84]. Definition (A.1) means that $f(\boldsymbol{X})/t$ is the Boolean function which is equivalent to $f(\boldsymbol{X})$ whenever $t$ is equal to 1, and hence it can be obtained by forming the conjunction $f(\boldsymbol{X}) \wedge t$ of $f(\boldsymbol{X})$ and $t$, and then suppressing $t$ from this conjunction [79]. The Boolean quotient $f(\boldsymbol{X})/t$ is a function of the variables $\boldsymbol{X}$ except for any variable that appears in $t$. Brown [44] lists and proves several useful properties of Boolean quotients, of which we reproduce the following ones:

$$f(\boldsymbol{X})/1 = f(\boldsymbol{X})/Y = f(\boldsymbol{X}), \qquad (A.2)$$

$$f(\boldsymbol{X})/st = (f(\boldsymbol{X})/s)/t = (f(\boldsymbol{X})/t)/s\,, \qquad \text{for } st \neq 0, \qquad (A.3)$$

$$f(\boldsymbol{X}) \leq g\,(\boldsymbol{X}) \;\Longrightarrow\; f(\boldsymbol{X})/t \;\leq\; g(\boldsymbol{X})/t, \qquad \text{for n-variable functions } f(\boldsymbol{X}) \text{ and } g(\boldsymbol{X}) \text{ and an m-variable term t with } m \leq n\,, \qquad (A.4)$$

$$t \wedge f(\boldsymbol{X}) = t \;\wedge (f(\boldsymbol{X})/t), \qquad (A.5)$$

$$\bar{t} \vee f(\boldsymbol{X}) = \bar{t} \vee \big(t \wedge f(\boldsymbol{X})\big) = \bar{t} \vee (t \wedge (f(\boldsymbol{X})/t)\,) = \bar{t} \vee (f(\boldsymbol{X})/t), \qquad (A.6)$$

$$t \;\wedge\; f(\boldsymbol{X}) \;\leq\; f(\boldsymbol{X})/t \;\;\leq\; \bar{t} \vee f(\boldsymbol{X}). \qquad (A.7)$$

If the term $t$ implies the function $f(\boldsymbol{X})$, i.e., if $t \;= t \;\wedge\; f(\boldsymbol{X})$, then [79]

$$f(\boldsymbol{X})/t = \; f(\boldsymbol{X})/(t \;\wedge\; f(\boldsymbol{X})) = 1. \qquad \text{(A.8)}$$

The formation of a Boolean quotient commutes with the negation (complementation), ANDing (∧), ORing (∨), and XORing (⊕) operations (on Boolean functions of the same arguments or of different arguments $\boldsymbol{X}$ and $\boldsymbol{Y}$), i.e.

$$\overline{f(\boldsymbol{X})/X_m} = \bar{f}(\boldsymbol{X})/X_m, \qquad \text{(A.9)}$$

$$(f_1(\boldsymbol{X}) \;\wedge\; f_2(\boldsymbol{Y}))/t \;=\; (f_1(\boldsymbol{X})/t) \;\wedge\; (f_2(\boldsymbol{Y})\,/t), \qquad \text{(A.10)}$$

$$(f_1(\boldsymbol{X}) \;\vee\; f_2(\boldsymbol{Y}))/t \;=\; (f_1(\boldsymbol{X})/t) \;\vee\; (f_2(\boldsymbol{Y})/t), \qquad \text{(A.11)}$$

$$(f_1(\boldsymbol{X}) \;\oplus\; f_2(\boldsymbol{Y}))/t \;=\; (f_1(\boldsymbol{X})/t) \;\oplus\; (f_2(\boldsymbol{Y})/t). \qquad \text{(A.12)}$$

In this Appendix, we followed Brown [44] in denoting a Boolean quotient by an inclined slash $(f/t)$. However, it is possible to denote it by a vertical bar $(f|t)$ to stress the equivalent meaning (borrowed from conditional probability) of $f$ conditioned by $t$ or $f$ given $t$. We hope that the inclined slash for the Boolean quotient $(f/t)$ does not get confused with the one in $(\boldsymbol{X}/X_m)$ which denotes the vector $\boldsymbol{X}$ with its $X_m$ component excluded.

Two concepts quite related to that of the Boolean quotient are that of the Boole-Shannon expansion [48, 50, 51, 54, 65], and the Boolean derivative (difference) [57, 58]. In fact, the Boole-Shannon expansion might be deduced in terms of the Boolean quotient as follows:

$$f(\boldsymbol{X}) = f(\boldsymbol{X}) \wedge (\bar{X}_m \vee X_m) \;= f(\boldsymbol{X})\,\bar{X}_m \;\vee\; f(\boldsymbol{X})\,X_m = (f(\boldsymbol{X})/\,\bar{X}_m)\;\bar{X}_m \;\vee\; (f(\boldsymbol{X})/X_m)\,X_m = (f(\boldsymbol{X})/\,\bar{X}_m)\;\bar{X}_m \;\oplus\; (f(\boldsymbol{X})/X_m)\,X_m. \qquad \text{(A.13)}$$

Note that it makes no difference to consider the Boole-Shannon expansion in (A.13) as the ORing (∨) or XORing (⊕) of the two terms $(f(\boldsymbol{X})/\,\bar{X}_m)\;\bar{X}_m$ and $(f(\boldsymbol{X})/X_m)\,X_m$, which are disjoint terms since the former contains the complemented literal $\bar{X}_m$, while the latter contains the un- complemented literal $X_m$. Likewise, the Boolean derivative (difference) might be deduced as the XORing of two Boolean quotients as follows [57, 58]

$$\frac{\partial f(\boldsymbol{X})}{\partial X_m} = (f(\boldsymbol{X})/X_m) \oplus (f(\boldsymbol{X})/\overline{X}_m). \qquad \text{(A.14)}$$

Now, we show that orthogonality is preserved when Boolean quotients are introduced. Starting with two orthogonal functions $f_1(\boldsymbol{X})$ and $f_2(\boldsymbol{X})$, i.e., with

$$f_1(\boldsymbol{X}) \ \wedge \ f_2(\boldsymbol{X}) = 0, \qquad \text{(A.15)}$$

we obtain via (A.13)

$$((f_1(\boldsymbol{X})/ \bar{X}_m) \wedge \bar{X}_m \ \vee \ (f_1(\boldsymbol{X})/X_m) \ \wedge X_m) \ \wedge \ (f_2(\boldsymbol{X})/ \bar{X}_m) \wedge \ \bar{X}_m \ \vee \ (f_2(\boldsymbol{X})/X_m) \wedge X_m) = 0.$$

$$(f_1(\boldsymbol{X})/ \bar{X}_m) \wedge (f_2(\boldsymbol{X})/ \bar{X}_m) \wedge \bar{X}_m \ \vee \ (f_1(\boldsymbol{X})/X_m) \wedge (f_2(\boldsymbol{X})/X_m) \wedge X_m = 0,$$

which can be shown to imply the orthogonality of the Boolean quotients of the original functions w.r.t arbitrary literals $\bar{X}_m$ or $X_m$.

$$(f_1(\boldsymbol{X})/ \bar{X}_m) \wedge (f_2(\boldsymbol{X})/ \bar{X}_m) = 0, \qquad \text{(A.16a)}$$

$$(f_1(\boldsymbol{X})/X_m) \wedge (f_2(\boldsymbol{X})/X_m) = 0. \qquad \text{(A.16b)}$$

A particular case of the above result (proved earlier in [12]) is that disjointness of a sum-of-products form is preserved by the construction of its Boolean quotient w.r.t. any pertinent term $t$. The sum-of-products (s-o-p) form $\vee_{k=1}^{m} D_k$ is disjoint if every two products $D_{k_1}$ and $D_{k_2}$ in it are disjoint (orthogonal), i.e., $D_{k_1} \wedge D_{k_2} = 0$. Hence, we can assert disjointness preservation if we substitute $D_{k_1}$ for $f_1(\boldsymbol{X})$ and $D_{k_2}$ for $f_2(\boldsymbol{X})$ in the above results in (A.16).

The concept of Boolean quotients can play a crucial role in the formulation of a switching-algebraic theory for voting systems. Boolean-quotient formulas (5) and (6) express the Banzhaf voting powers for general voting systems that are not necessarily monotone, while Boolean-quotient formulas (10)-(15) express the Banzhaf voting powers for monotone voting systems. Many fundamental concepts of voting theory have Boolean-quotient interpretations, as we now show.

A voter $X_m$ possesses *veto* power [9, 12, 85] if his/her support is necessary for the quota to be reached, and hence he/she is critical in every winning coalition. This happens when $\overline{X}_m$ implies $\overline{f}(\boldsymbol{X})$, i.e., $\overline{X}_m f(\boldsymbol{X}) = 0$ or $f(\boldsymbol{X})/\overline{X}_m = 0$ $(\overline{f}(\boldsymbol{X})/\overline{X}_m = 1)$. In this case

$$f(\boldsymbol{X}) = X_m \ (f(\boldsymbol{X})/X_m) = X_m \ g(\boldsymbol{X}/X_m). \qquad \text{(A.17)}$$

As a notable example of veto power, the decision function for the voting system of the United Nations Security Council (UNSC) (with no members absent or abstaining) is given by $f(\boldsymbol{P};\boldsymbol{N}) = P_1 \ P_2 \ P_3 \ P_4 \ P_5 \ Sy(10; \{4..10\}, \boldsymbol{N})$, wherein each of the five permanent members $P_i$ enjoys veto power.

A voter $X_m$ is a *dummy* voter [9,12, 21, 28, 86] iff he/she is not critical in any winning coalition, and hence the decision function $f(\boldsymbol{X})$ is independent of $X_m$ ($\frac{\partial f(\boldsymbol{X})}{\partial X_m} = 0$), i.e.,

$$f(\boldsymbol{X})/X_m = f(\boldsymbol{X})/\overline{X}_m. \tag{A.18}$$

A voter $X_m$ is a *dictator* [9, 12, 19, 87] iff his/her weight is equal to or greater than the quota (and hence strictly greater than half the sum of weights). This means that the single-voter coalition $\{X_m\}$ is winning. A dictator has veto power while all other voters are dummies, so that the system decision is exactly his/hers ($f(\boldsymbol{X}) = X_m$). In this case

$$f(\boldsymbol{X})/X_m = 1 \neq f(\boldsymbol{X})/\overline{X}_m = 0, \quad f(\boldsymbol{X})/X_i = f(\boldsymbol{X})/\overline{X}_i \quad \forall \;\; i \neq m. \tag{A.19}$$

Two voters $X_i$ and $X_j$ constitute a *clique* [19, 88] ($f(\boldsymbol{X}) = X_i X_j$) iff each of them has veto power and all other voters are dummies, i.e.,

$$f(\boldsymbol{X})/X_i X_j = 1 \neq f(\boldsymbol{X})/\overline{X_i X}_j = 0, \quad f(\boldsymbol{X})/X_k = f(\boldsymbol{X})/\overline{X}_k \quad \forall \;\; k \neq i, k \neq j. \tag{A.20}$$

Similarly, we can define a clique of three or more voters. If we assume $Sy(10; \{4..10\}, \boldsymbol{N}) = 1$ in the decision function for the UNSC voting system, then the five permanent members constitute a clique.

A distinct advantage of the Boolean quotient in voting theory is its utility in interpreting the concept of 'desirability' of a voter in terms of the formation of a winning coalition [6, 9, 89]. To decide whether a voter $X_i$ is comparable or not to another $X_j$, we compute the Boolean quotient $A_i = f(\boldsymbol{X})/X_i$, which is a function of $\boldsymbol{X}/X_i$. We also compute the Boolean quotient $f(\boldsymbol{X})/X_j$, which is a function of $\boldsymbol{X}/X_j$, and then replace each $X_i$ in it by $X_j$ to obtain $A_j = \left(f(\boldsymbol{X})/X_j\right)_{X_i \to X_j}$, which is now a function of $\boldsymbol{X}/X_i$. The two voters $X_i$ and $X_j$ are equally desirable (or equivalent) iff the two switching functions $A_i$ and $A_j$ are equal ($A_i = A_j \; or \; (A_i \geq A_j) \cap (A_i \leq A_j)$). This happens when the decision function $f(\boldsymbol{X})$ is partially symmetric in $X_i$ and $X_j$. The voter $X_i$ is more desirable than the voter $X_j$ iff $A_i \geq A_j$, and vice versa. The two voters $X_i$ and $X_j$ are incomparable if neither $A_i \geq A_j$ nor $A_i \leq A_j$. As a quick example, we consider the bi-cameral (vector-weighted) voting system:

$$f(\boldsymbol{X}) = (X_1 \;\vee\; X_2)\;(X_3 \;\vee\; X_4 X_5). \tag{A.21}$$

Here, the two voters $X_1$ and $X_2$ are equally desirable (or equivalent), and so are the two voters $X_4$ and $X_5$. The voter $X_3$ is more desirable than each of the voters $X_4$ and $X_5$. Each of the two voters $X_1$ and $X_2$ (of the first chamber) is incomparable to each of the three voters $X_3, X_4$ and $X_5$ (of the second chamber).

Next, we consider a sophisticated yes-no voting system, namely the US federal voting system. In this system, the Boolean (or Bernoulli) variables $P, V, S_k$ and $H_j$ represent the voting of the President, Vice-President, senator $k$ (member $k$ of the Senate) $(1 \leq k \leq 100)$, and member $j$ of the House of Representatives $(1 \leq j \leq 435)$, which amount to a total of $537$ distinct individual voters in the system. The decision function $f = f(P, V, \boldsymbol{S}, \boldsymbol{H})$ of this system is [30]

$$f = \bar{P}\ Sy(100;\ \{67..100\}, \boldsymbol{S})\ \ Sy(435;\ \{290..435\}, \boldsymbol{H}) \vee$$
$$P\, \bar{V}\ Sy(100;\ \{51..100\}, \boldsymbol{S})\ \ Sy(435;\ \{218..435\}, \boldsymbol{H}) \vee$$
$$P\, V\, Sy(100;\ \{50..100\}, \boldsymbol{S})\ \ Sy(435;\ \{218..435\}, \boldsymbol{H}). \quad \text{(A.22)}$$

Relations of desirability among members of this sophisticated US federal voting system can be satisfactorily obtained via the Boolean-quotient machinery. A notable example of these relations (that might be anticipated in an intuitionistic fashion) is that a senator $S_k$ and a member of the House of Representatives $H_j$ are incomparable [9]. The President $P$ is more desirable than any other individual voter. The Vice-President $V$ is incomparable to a member of the House of Representatives, and less desirable than any senator $S_k$ [9]. We now employ the above Boolean-quotient methodology to prove this latter relation. Specifically, we employ switching-algebraic techniques to calculate:

$$(f/\,S_k)_{V \to S_k} = \bar{P}\ Sy(99;\ \{66..99\}, \boldsymbol{S}/\,S_k)\ \ Sy(435;\ \{290..435\}, \boldsymbol{H}) \vee$$
$$P\, \overline{S_k}\ Sy(99;\ \{50..99\}, \boldsymbol{S}/\,S_k)\ \ Sy(435;\ \{218..435\}, \boldsymbol{H}) \vee$$
$$P\ S_k\, Sy(99;\ \{49..99\}, \boldsymbol{S}/\,S_k)\ \ Sy(435;\ \{218..435\}, \boldsymbol{H})$$
$$= \bar{P}\ Sy(99;\ \{66..99\}, \boldsymbol{S}/\,S_k)\ \ Sy(435;\ \{290..435\}, \boldsymbol{H}) \vee$$
$$P\, Sy(100;\ \{50..100\}, \boldsymbol{S})\ \ Sy(435;\ \{218..435\}, \boldsymbol{H}). \quad \text{(A.23)}$$

$$f/V = \bar{P}\ Sy(100;\ \{67..100\}, \boldsymbol{S})\ \ Sy(435;\ \{290..435\}, \boldsymbol{H}) \vee$$
$$P\ Sy(100;\ \{50..100\}, \boldsymbol{S})\ \ Sy(435;\ \{218..435\}, \boldsymbol{H}). \quad \text{(A.24)}$$

We note that $f/V$ in (A.24) and $(f/\,S_k)_{V \to S_k}$ in (A.23) are exactly the same with the sole exception of the appearance of the symmetric switching function (SSF) $Sy(100;\ \{67..100\}, \boldsymbol{S})$ in the former and the SSF $Sy(99;\ \{66..99\}, \boldsymbol{S}/\,S_k)$ in the latter. These two SSFs satisfy

$$Sy(100;\ \{67..100\}, \boldsymbol{S}) = \overline{S_k}\ Sy(99;\ \{67..99\}, \boldsymbol{S}/\,S_k) \vee S_k\, Sy(99;\ \{66..99\}, \boldsymbol{S}/\,S_k) \leq$$
$$\overline{S_k}\ Sy(99;\ \{66..99\}, \boldsymbol{S}/\,S_k) \vee S_k\, Sy(99;\ \{66..99\}, \boldsymbol{S}/\,S_k) = Sy(99;\ \{66..99\}, \boldsymbol{S}/\,S_k), \quad \text{(A.25)}$$

which automatically leads to the desired result that the Vice-President $V$ is less desirable than any senator $S_k$, namely

$$f/V \leq (f/S_k)_{V\rightarrow S_k}. \tag{A.26}$$

In contrast to the Vice-President, the President $P$ is more desirable than any senator $S_k$. To prove this, we compute

$$(f/S_k)_{P\rightarrow S_k} = \overline{S_k}\ Sy(99;\ \{66..99\}, \boldsymbol{S}/S_k)\ \ Sy(435;\ \{290..435\}, \boldsymbol{H}) \vee$$

$$\bar{V}\ S_k\ Sy(99;\ \{50..99\}, \boldsymbol{S}/S_k)\ \ Sy(435;\ \{218..435\}, \boldsymbol{H}) \vee$$

$$V\ S_k\, Sy(99;\ \{49..99\}, \boldsymbol{S}/S_k)\ \ Sy(435;\ \{218..435\}, \boldsymbol{H}). \tag{A.27}$$

$$f/P = \bar{V}\ Sy(100;\ \{51..100\}, \boldsymbol{S})\ \ Sy(435;\ \{218..435\}, \boldsymbol{H}) \vee$$

$$V\, Sy(100;\ \{50..100\}, \boldsymbol{S})\ \ Sy(435;\ \{218..435\}, \boldsymbol{H}). \tag{A.28}$$

Thanks to the inequalities

$$S_k\ Sy(99;\ \{50..99\}, \boldsymbol{S}/S_k) \leq Sy(100;\ \{51..100\}, \boldsymbol{S}),$$

$$S_k\, Sy(99;\ \{49..99\}, \boldsymbol{S}/S_k) \leq Sy(100;\ \{51..100\}, \boldsymbol{S}).$$

we obtain the required result, namely

$$(f/S_k)_{P\rightarrow S_k} \leq f/P. \tag{A.29}$$

Due to transitivity of desirability, we expect (A.26) and (A.29) to combine to assert that the President is more desirable than the Vice-President, a fact that we prove by computing

$$(f/P)_{V\rightarrow P} = \bar{P}\ Sy(100;\ \{51..100\}, \boldsymbol{S})\ \ Sy(435;\ \{218..435\}, \boldsymbol{H}) \vee$$

$$P\, Sy(100;\ \{50..100\}, \boldsymbol{S})\ \ Sy(435;\ \{218..435\}, \boldsymbol{H}). \tag{A.30}$$

We note that $f/V$ in (A.24) and $(f/P)_{V\rightarrow P}$ in (A. 30) are exactly the same with the sole exception of the appearance of the $Sy(100;\ \{67..100\}, \boldsymbol{S})$ $Sy(435;\ \{290..435\}, \boldsymbol{H})$ SSF in the former and the SSF $Sy(100;\ \{51..100\}, \boldsymbol{S})$ $Sy(435;\ \{218..435\}, \boldsymbol{H})$ in the latter. These two SSFs obviously satisfy

$$Sy(100;\ \{67..100\}, \boldsymbol{S})\, Sy(435;\ \{290..435\}, \boldsymbol{H}) \leq Sy(100;\ \{51..100\}, \boldsymbol{S})\ Sy(435;\ \{218..435\}, \boldsymbol{H}).$$

and hence, we obtain the required result.

$$f/V \leq (f/P)_{V\rightarrow P}. \tag{A.31}$$

## Appendix B: The k-out-of-n switching functions

The k-out-of-n function $(0 \leq k \leq n)$ is a monotonically non-decreasing symmetric switching function of characteristic set $\{m|\ k \leq m \leq n\} = \{k..n\}$, and hence is denoted $Sy(n;\{k..n\};\boldsymbol{X})$ [12, 50, 90-94]. It can also be viewed as a symmetric coherent threshold function with unit weights and a threshold equal to $k$, since

$$\{Sy(n;\{k..n\};\boldsymbol{X}) = 1\} \quad iff \quad \{\sum_{i=1}^{n} X_i \geq k\}. \tag{B.1}$$

The celebrated simple majority voting corresponds to $k$ being equal to the ceiling of $(n+1)/2$. No lower value of $k$ should be permitted practically due to the requirement that $k$ not allowed to be less than half the sum $n$ of weights. Higher values of $k$ allow modeling the systems of super-majority such as the ones with quotas of two thirds or four fifths. Note that in (B.1) the symbol $n$ retains its natural role in designating the number of variables (components or elements), while the symbol $k$ depicts the threshold or quota, as expected. This is contrary to widespread usage [18, 20, 24, 26, 95, 96], wherein a threshold system is inadvertently renamed as a weighted k-out-of-n system, and the symbol $k$ assumes the role of the number of components while the symbol $n$ deviates from its customary meaning to usurp the role of the threshold. With this unfortunate renaming, you can hear of the ridiculous name of a weighted 9-out-of-5 system, which defies the streamlined logic of the English language.

The Boole-Shannon expansion for the k-out-of-n function about any of its variables $X_m$ $(1 \leq m \leq n)$ can be stated as follows [90-94]

$$Sy(n;\{k..n\};\boldsymbol{X}) = \bar{X}_m\ Sy(n-1;\{k..(n-1)\};\boldsymbol{X}/X_m) \ \vee \ X_m\ Sy(n-1;\{(k-1)..(n-1)\};\boldsymbol{X}/X_m), \ (1 \leq m \leq n), \tag{B.2}$$

The expansion (B.2) can be recursively applied till one of the following boundary conditions is reached:

$$Sy(n;\{0..n\};\boldsymbol{X}) = 1, \tag{B.3}$$

$$Sy(n;\boldsymbol{\phi};\boldsymbol{X}) = 0, \tag{B.4}$$

where $\{0..n\} = \boldsymbol{I}_{n+1} = \{0,1,2, \ldots, n\}$ is the universe of discourse for the first $(n+1)$ non-negative integers, and $\boldsymbol{\phi} = \{\}$ is the empty set (null set or set with no elements).

The two terms in the RHS of (B.2) are disjoint since $\bar{X}_m$ appears in the first term while $X_m$ appears in the second. Therefore, it is legitimate to replace the OR operator ($\vee$) by an XOR operator ($\oplus$) in (B.2), namely

$$Sy(n;\{k..n\};\boldsymbol{X}) = \bar{X}_m\ Sy(n-1;\{k..(n-1)\};\boldsymbol{X}/X_m) \ \oplus \ X_m\ Sy(n-1;\{(k-1)..(n-1)\};\boldsymbol{X}/X_m), \ (1 \leq m \leq n), \tag{B.5}$$

Hence, the Boolean derivative of the k-out-of-n function w.r.t. $X_m$ is readily obtained as another symmetric switching function given by

$$\frac{\partial Sy(n;\{k..n\};X)}{\partial X_m} = Sy(n-1;\{k..(n-1)\};\boldsymbol{X}/X_m) \ \oplus \ Sy(n-1;\{(k-1)..(n-1)\};\boldsymbol{X}/X_m), \quad (1 \leq m \leq n), \tag{B.6}$$

$$\frac{\partial Sy(n;\{k..n\};X)}{\partial X_m} = Sy(n-1;\{k-1\};\boldsymbol{X}/X_m), \quad (1 \leq m \leq n). \tag{B.7}$$

The total Banzhaf power for each element of a k-out-of-n voting system is given by the weight of this function, namely

$$TBP(X_m) = c(n-1, k-1), \quad (1 \leq m \leq n), \tag{B.8}$$

where the function $c(n,k)$ denotes the binomial (combinatorial) coefficient, labelled as $n\, choose\, k$ , or the number of ways of choosing $k$ out of $n$ objects when repetition is not allowed, and order does not matter. For comparison, we note that the weight of the k-out-of-n function $(0 \leq k \leq n)$ is given by

$$wt\big(Sy(n;\{k..n\};\boldsymbol{X})\big) = C(n,k) = \textstyle\sum_{m=k}^{n} c(n,m), \tag{B.9}$$

where the upper–case $C(n,k)$ denotes the cumulative combinatorial coefficient [12]. The normalized total Banzhaf power for each element of a k-out-of-n voting system is given by

$$NTBP(X_m) = \frac{1}{n}, \quad (1 \leq m \leq n), \tag{B.10}$$

In retrospect, we note that we might not have really needed to carry out the aforementioned detailed calculations, because we could have deduced directly from the symmetry of the k-out-of-n voting system that the power of the voters is going to be equal.

Since the function $Sy(n;\{k..n\};\boldsymbol{X})$ is monotonically non-decreasing, its 0-subfunction implies its 1-subfunction, i.e.,

$$Sy(n-1;\{k..(n-1)\};\boldsymbol{X}/X_m) \leq Sy(n-1;\{(k-1)..(n-1)\};\boldsymbol{X}/X_m), \tag{B.11}$$

and hence it can be rewritten as

$$Sy(n;\{k..n\};\boldsymbol{X}) = Sy(n-1;\{k..(n-1)\};\boldsymbol{X}/X_m) \vee X_m\, Sy(n-1;\{(k-1)..(n-1)\};\boldsymbol{X}/X_m), \quad (1 \leq m \leq n), \tag{B.12}$$

Note that $X_m$ appears solely as un-complemented in (B.12). The public good index (PGI) is the number of prime implicants in which $X_m$ appears, and hence it is the number of prime implicants of the $(k-1)$-out-of-$(n-1)$ function $Sy(n-1;\{(k-1)..(n-1)\};\boldsymbol{X}/X_m)$, which is the binomial coefficient. $c(n-1,k-1)$. This result means that $PGI(X_m)$ is exactly equal to $TBP(X_m)$ for an unrestricted k-out-of-n system.

## References

[ 1] Dodd, L. C. (1974). Party coalitions in multiparty parliaments: A game-theoretic analysis. *American Political Science Review*, *68*(3), 1093-1117.

[ 2] Palfrey, T. R., & Rosenthal, H. (1983). A strategic calculus of voting. *Public choice*, *41*(1), 7-53.

[ 3] Staatz, J. M. (1983). The cooperative as a coalition: a game-theoretic approach. *American Journal of Agricultural Economics*, *65*(5), 1084-1089.
[ 4] Taylor, A., & Zwicker, W. (1992). A characterization of weighted voting. *Proceedings of the American mathematical society*, *115*(4), 1089-1094.
[ 5] Gates, S., & Humes, B. D. (1997). *Games, Information, and Politics: Applying Game Theoretic Models to Political Science*. University of Michigan Press.
[ 6] Taylor, A. D., & Zwicker, W. S. (2000). *Simple Games: Desirability Relations, Trading, Pseudoweightings*. Princeton University Press.
[ 7] Peleg, B. (2002). Game-theoretic analysis of voting in committees. *Handbook of social choice and welfare*, *1*, 395-423.
[ 8] Krueger, J. I., & Acevedo, M. (2008). A game-theoretic view of voting. *Journal of Social Issues*, *64*(3), 467-485.
[ 9] Taylor, A. D., & Pacelli, A. M. (2008). *Mathematics and Politics: Strategy, Voting, Power, and Proof,* 2nd Edition, Springer Science & Business Media, New York, NY, USA.
[ 10] Maaser, N. F. (2010). *Decision-Making in Committees: Game-Theoretic Analysis* (Vol. 635). Springer Science & Business Media.
[ 11] Liu, Z., Luong, N. C., Wang, W., Niyato, D., Wang, P., Liang, Y. C., & Kim, D. I. (2019). A survey on blockchain: A game theoretical perspective. *IEEE Access*, *7*, 47615-47643.
[ 12] Rushdi, A. M. & Rushdi, M. A. (2023). Towards a switching-algebraic theory of weighted monotone voting systems: The case of Banzhaf voting indices, a*rXiv Preprint*, 1-34, https://doi.org/10.48550/arXiv.2302.09367.
[ 13] Rushdi, A. M. (1990). Threshold systems and their reliability. *Microelectronics and Reliability*, *30*(2), 299-312.
[ 14] O'Donnell, R. (2008, May). Some topics in analysis of Boolean functions. In *Proceedings of the Fortieth Annual ACM Symposium on Theory of Computing* (pp. 569-578).
[ 15] Yamamoto, Y. (2012, May). Banzhaf index and Boolean difference. In *2012 IEEE 42nd International Symposium on Multiple-Valued Logic* (pp. 191-196). IEEE.
[ 16] Yamamoto, Y. (2014).Power Indices and Logic Functions. *Journal of Takasaki City University of Economics*, 56(4) 1-16 (in Japanese).
[ 17] O'Donnell, R. (2014). Social choice, computational complexity, Gaussian geometry, and Boolean functions, a*rXiv Preprint*, 1-27, https://arxiv.org/pdf/1407.7763.pdf.
[ 18] Rushdi, A. M. A., & Alturki, A. M. (2015). Reliability of coherent threshold systems. *Journal of Applied Sciences*, *15*(3), 431-443.
[ 19] Alturki, A. M., & Rushdi, A. M. A. (2016). Weighted voting systems: A threshold-Boolean perspective. *Journal of Engineering Research*, *4*, 1-19.
[ 20] Rushdi, A. M. A., & Bjaili, H. A. (2016). An ROBDD algorithm for the reliability of double-threshold systems. *British Journal of Mathematics and Computer Science*, *19*(6), 1-17.
[ 21] Rushdi, A. M. A., & Ba-Rukab, O. M. (2017). Calculation of Banzhaf voting indices utilizing variable-entered Karnaugh maps. *Journal of Advances in Mathematics and Computer Science*, *20*(4), 1-17.
[ 22] Rushdi, A. M. A., & Ba-Rukab, O. M. (2017). Map calculation of the Shapley-Shubik voting powers: An example of the European Economic Community. *International Journal of Mathematical, Engineering and Management Sciences*, *2*(1), 17-29.

[ 23] Rushdi, A. M. A., & Alturki, A. M. (2017). An application of reliability-analysis techniques in project management. *Journal of Advances in Mathematics and Computer Science*, *21*(6), 1-15.
[ 24] Rushdi, A. M. A., & Alturki, A. M. (2017). Computation of k-out-of-n system reliability via reduced ordered binary decision diagrams. *British Journal of Mathematics & Computer Science*, *22*(3), 1-9.
[ 25] Rushdi, R. A., & Rushdi, A. M. (2018). Karnaugh-map utility in medical studies: The case of Fetal Malnutrition. *International Journal of Mathematical, Engineering and Management Sciences*, *3*(3), 220-244.
[ 26] Rushdi, A. M., & Alturki, A. M. (2018). Novel representations for a coherent threshold reliability system: a tale of eight signal flow graphs. *Turkish Journal of Electrical Engineering and Computer Sciences*, *26*(1), 257-269.
[ 27] Freixas, J. (2021). On the enumeration of Boolean functions with distinguished variables. *Soft Computing*, 25(19), 12627-12640.
[ 28] Rushdi, A. M. A., & Ba-Rukab, O. M. (2019). Translation of Weighted Voting Concepts to the Boolean Domain: The Case of the Banzhaf Index. Chapter 10 in *Advances in Mathematics and Computer Science Vol. 2*, Book Publisher International, Hooghly, West Bengal, India, 122-140.
[ 29] Rushdi, A. M., & Rushdi, M. A. (2023). Switching-algebraic calculation of Banzhaf voting indices, *arXiv Preprint*, 1-18. https://doi.org/10.48550/arXiv.2302.09367. Also in *Journal of Computational and Cognitive Engineering* (*JCCE*), 2(3).
[ 30] Rushdi, A. M., & Rushdi, M. A. (2023). Boolean-based probability: the case of a vector-weighted voting system. *King Abdulaziz University Journal: Engineering Sciences,34*(2).
[ 31] Derks, J., & Peters, H. (1993). A Shapley value for games with restricted coalitions. *International Journal of Game Theory*, *21*, 351-360.
[ 32] Alonso-Meijide, J. M., & Casas-Méndez, B. (2007). The Public Good Index when some voters are incompatible. *Homo Oeconomicus*, 24(3–4), 449-468.
[ 33] Yakuba, V. (2008). Evaluation of Banzhaf index with restrictions on coalitions formation. *Mathematical and Computer Modelling*, *48*(9-10), 1602-1610.
[ 34] Tanino, T. (2009). Multiobjective cooperative games with restrictions on coalitions. In *Multiobjective Programming and Goal Programming: Theoretical Results and Practical Applications* (pp. 167-174). Springer Berlin Heidelberg.
[ 35] Shvarts, D. A. (2009). On calculation of the power indices with allowance for the agent preferences. *Automation and Remote Control*, 70, 484-490.
[ 36] Alonso-Meijide, J. M., Alvares-Mozos, M., & Fiestras-Janeiro, M. G. (2009). The Banzhaf value when some players are incompatible. *Homo Oeconomicus*, *26*(3/4), 403-414.
[ 37] Alonso-Meijide, J. M., Casas-Méndez, B., & Fiestras-Janeiro, M. G. (2013). A review of some recent results on power indices. In *Power, Voting, and Voting Power: 30 Years after*, 231-245.
[ 38] Alonso-Meijide, J. M., Casas-Méndez, B., & Fiestras-Janeiro, M. G. (2015). Computing Banzhaf–Coleman and Shapley–Shubik power indices with incompatible players. *Applied Mathematics and Computation*, *252*, 377-387.
[ 39] Alonso-Meijide, J. M., Casas-Méndez, B., & Fiestras-Janeiro, M. G. (2015). Computing Banzhaf–Coleman and Shapley–Shubik power indices with incompatible players. *Applied Mathematics and Computation*, *252*, 377-387.
[ 40] Rodríguez-Veiga, J., Novoa-Flores, G. I., & Casas-Méndez, B. (2016). Implementing generating functions to obtain power indices with coalition configuration. *Discrete Applied Mathematics*, *214*, 1-15.

[ 41] Francisco Neto, A., & Fonseca, C. R. (2019). An approach via generating functions to compute power indices of multiple weighted voting games with incompatible players. *Annals of Operations Research*, 279, 221-249.
[ 42] Skibski, O., Suzuki, T., Grabowski, T., Sakurai, Y., Michalak, T., & Yokoo, M. (2022). Measuring power in coalitional games with friends, enemies and allies. *Artificial Intelligence*, *313*, 103792.
[ 43] Bilbao, J. M., Fernandez, J. R., Losada, A. J., & Lopez, J. J. (2000). Generating functions for computing power indices efficiently. *Top*, *8*(2), 191-213.
[ 44] Brown, F. M. (1990). *Boolean Reasoning: The Logic of Boolean Equations*, Kluwer Academic Publishers, Boston, MA, USA.
[ 45] Rushdi, A. M., & Ba-Rukab, O. M. (2009). An exposition of the modern syllogistic method of propositional logic. *Journal of Umm Al-Qura University: Engineering and Architecture*, *1*(1), 17-49.
[ 46] Rushdi, A. M. A., & Zagzoog, S. S. (2018). Design of a digital circuit for integer factorization via solving the inverse problem of logic. *Journal of Advances in Mathematics and Computer Science*, *26*(3), 1-14.
[ 47] Rushdi, A. M., and Rushdi, M. A. (2016, November). Switching-algebraic algorithmic derivation of candidate keys in relational databases. In *2016 International Conference on Emerging Trends in Communication Technologies (ETCT)* (pp. 1-5). IEEE.
[ 48] Rushdi, A. M. and Rushdi, M. A. (2017). Switching-Algebraic Analysis of System Reliability. Chapter 6 in M. Ram and P. Davim (Editors), *Advances in Reliability and System Engineering*, Management and Industrial Engineering Series, Springer International Publishing, Cham, Switzerland, 139-161.
[ 49] Rushdi, A. M., & Rushdi, M. A. (2018). Mathematics and Examples of the Modern Syllogistic Method of Propositional Logic. Chapter 6 in M. Ram (Editor), *Mathematics Applied in Information Systems*, Bentham Science Publishers, Emirate of Sharjah, United Arab Emirates, 6, 123-167.
[ 50] Rushdi, A. M. A., (2019). Utilization of symmetric switching functions in the symbolic reliability analysis of multi-state k-out-of-n systems. *International Journal of Mathematical, Engineering and Management Sciences*, 4(2), 306-326.
[ 51] Rushdi, A. M., & Amashah, M. H. (2021). A liaison among inclusion-exclusion, probability ready expressions and Boole-Shannon expansion for multi-state reliability. *Journal of King Abdulaziz University: Computing and Information Technology Sciences*, *10*(2), 1-17.
[ 52] Rushdi, A. M. A., & Badawi, R. S. (2021). Map Visualization of Cause-Effect Relations in Qualitative Comparative Analysis. Chapter 1 in *Novel Perspectives of Engineering Research Vol. 3*, Book Publisher International, Hooghly, West Bengal, India 1-24.
[ 53] Rushdi, A. M. & D. L. Al-Khateeb. (1983). A review of methods for system reliability analysis: A Karnaugh-map perspective, *Proceedings of the First Saudi Engineering Conference*, Jeddah, Saudi Arabia, Vol. 1, pp. 57-95.
[ 54] Rushdi, A. M. A., & Hassan, A. K. (2016). An exposition of system reliability analysis with an ecological perspective. *Ecological Indicators*, *63*, 282-295.
[ 55] Rushdi, A. M. A. & Amashah, M. H. (2021). Conventional and improved inclusion-exclusion derivations of symbolic expressions for the reliability of a multi-state network, *Asian Journal of Research in Computer Science*, *8*(1): 21-45.
[ 56] Rushdi, A. M. A., & Amashah, M. H. (2023).The Multi-State Inclusion-Exclusion Principle: Conventional and Improved Versions, Chapter 9 in *Research Highlights in Mathematics and Computer Science Vol. 5*, Book Publisher International, Hooghly, West Bengal, India, 87-128.

[ 57] Lee, S. C. (1978). *Modern Switching Theory and Digital Design*, Prentice-Hall, Englewood Cliffs, New Jersey, NJ, USA.

[ 58] Rushdi, A. M. (1986). Map differentiation of switching functions, *Microelectronics and Reliability*, 26(5), 891-907.

[ 59] Abraham, J. A. (1979). An improved algorithm for network reliability, *IEEE Transactions on Reliability,* R-28 (1), 58-61.

[ 60] Dotson, W., and Gobien J. (1979). A new analysis technique for probabilistic graphs, *IEEE Transactions on Circuits and Systems,* CAS-26(10): 855-865.

[ 61] Savir, J. (1980). Syndrome-testable design of combinational circuits. IEEE Transactions on Computers, 29(06), 442-451.

[ 62] Bennetts, R. G. (1982). Analysis of reliability block diagrams by Boolean techniques, *IEEE Transactions on Reliability,* R-31(2), 159-166.

[ 63] Rushdi, A. M. (1987). On computing the syndrome of a switching function, *Microelectronics and Reliability*, 27(4), 703-716.

[ 64] Rushdi, A. M. (1987). On computing the spectral coefficients of a switching function, *Microelectronics and Reliability*, 27(6), 965-979.

[ 65] Rushdi, A. M. A. & Ghaleb, F. A. M. (2015). The Walsh spectrum and the real transform of a switching function: A review with a Karnaugh-map perspective, *Journal of Qassim University: Engineering and Computer Sciences,* **7**(2), 73-112.

[ 66] Heard, A. D., and Swartz, T. B. (1999). Extended voting measures. *Canadian Journal of Statistics*, 27, 177–186.

[ 67] Ferger, W. F. (1931). The nature and use of the harmonic mean. *Journal of the American Statistical Association*, *26*(173), 36-40.

[ 68] Straffin, P. D. (1978). Probability models for power indices. In Game Theory and Political Science, ed. P. Ordeshook. New York University Press, 477–510.

[ 69] Straffin, P. D., Davis, M. D., & Brams, S. J. (1982). Power and satisfaction in an ideologically divided voting body. In *Power, voting, and voting power* (pp. 239-255). Physica-Verlag HD.

[ 70] Gelman, A., Katz, J. N., & Tuerlinckx, F. (2002). The mathematics and statistics of voting power. *Statistical Science*, 420-435.

[ 71] Kirsch, W. (2016). A mathematical view on voting and power. *Mathematics and Society, European Mathematical Society*, 251-279.

[ 72] O'neill, B. (1996). Power and satisfaction in the United Nations security council. *Journal of Conflict Resolution*, *40*(2), 219-237.

[ 73] Aczél, J., & Wagner, C. (1980). A characterization of weighted arithmetic means. *SIAM Journal on Algebraic Discrete Methods*, *1*(3), 259-260.

[ 74] Napel, S., & Widgrén, M. (2001). Inferior players in simple games. *International Journal of Game Theory*, *30*, 209-220.

[ 75] Holler, M. J. (1982). Forming coalitions and measuring voting power. *Political Studies*, 30(2), 262-271.

[ 76] Holler, M. J., & Packel, E. W. (1983). Power, luck and the right index. *Zeitschrift für Nationalökonomie/Journal of Economics*, *43*(1), 21-29.

[ 77] Holler, M. J. (2019). The story of the poor public good index. *Transactions on Computational Collective Intelligence XXXIV*, 171-179.

[ 78] Denver, D. (2007). ‘A historic moment’? The results of the Scottish Parliament elections 2007. Scottish Affairs, 60(1), 61-79.

[ 79] Ghazala, M. J. (1957). Irredundant disjunctive and conjunctive forms of a Boolean function. *IBM Journal of Research and Development*, *1*(2), 171-176.

[ 80] Majithia, J. C. (1972). A simple technique for determination of essential multiple output prime implicants. *IEEE Transactions on Computers*, C-21(9), 1024-1026.

[ 81] Reusch, B. (1975). Generation of prime implicants from subfunctions and a unifying approach to the covering problem. *IEEE Transactions on Computers*, C-24(9), 924-930.

[ 82] Rushdi, A. M. (2001). Prime-implicant extraction with the aid of the variable-entered Karnaugh map. *Umm Al-Qura University Journal: Science, Medicine and Engineering*, *13*(1), 53-74.

[ 83] Rushdi, A. M. A., & Badawi, R. M. S. (2017). Karnaugh-map utilization in Boolean analysis: The case of war termination. *Journal of Qassim University: Engineering and Computer Sciences*, *10*(1), 53-88.

[ 84] Bryant, R. E. (1986). Graph-based algorithms for Boolean function manipulation. *IEEE Transactions on Computers*, C-35(8), 677-691.

[ 85] Mashechkin, A. I., & Pospelova, I. I. (2009). Properties of voting with veto power. *Computational Mathematics and Modeling*, *20*(4), 427-437.

[ 86] Lucas, W. F. (1983). *Measuring power in weighted voting systems* (pp. 183-238). Springer New York.

[ 87] Saari, D. G. (2008). Disposing dictators, demystifying voting paradoxes. *Mathematical and Computer Modelling*, *48*(9-10), 1671-1673.

[ 88] Berg, D. E., Norine, S., Su, F. E., Thomas, R., & Wollan, P. (2010). Voting in agreeable societies. *The American Mathematical Monthly*, *117*(1), 27-39.

[ 89] Yamazaki, A., Inohara, T., & Nakano, B. (2000). Comparability of coalitions in committees with permission of voters by using desirability relation and hopefulness relation. *Applied Mathematics and computation*, *113*(2-3), 219-234.

[ 90] Rushdi, A. M. (1986). Utilization of symmetric switching functions in the computation of k-out-of-n system reliability, *Microelectronics and Reliability*, *26*(5): 973-987.

[ 91] Rushdi, A. M. (1991). Comments on "An efficient nonrecursive algorithm for computing the reliability of k-out-of-n systems." *IEEE Transactions on Reliability*, *40*(1), 60-61.

[ 92] Rushdi, A. M. (1993). Reliability of k-out-of-n Systems, Chapter 5 in Misra, K. B. (Editor), *New Trends in System Reliability Evaluation*, Vol. 16, Fundamental Studies in Engineering, Elsevier Science Publishers, Amsterdam, The Netherlands, 185-227.

[ 93] Rushdi, M. A. M., Ba-Rukab, O. M., & Rushdi, A. M. (2016). Multi-dimensional recursion relations and mathematical induction techniques: The case of failure frequency of k-out-of-n systems. *Journal of King Abdulaziz University: Engineering Sciences*, *27*(2), 15-31.

[ 94] Rushdi, A. M., & Alturki, A. M. (2018). Unification of mathematical concepts and algorithms of k-out-of-n system reliability: A perspective of improved disjoint products. *Journal of Engineering Research*, *6*(4), 1-31.

[ 95] Uswarman, R., & Rushdi, A. M. (2021). Reliability evaluation of rooftop solar photovoltaic using coherent threshold systems. *Journal of Engineering Research and Reports*, *20*(2), 32-44.

[ 96] Triantafyllou, I. S. (2023). Reliability structures consisting of weighted components: Synopsis and new advances. *Journal of Reliability and Statistical Studies*,*16*(1) 25-56.